\documentclass[iop]{emulateapj}
\usepackage{amsmath}

\usepackage{color}

\usepackage{lineno}
\shorttitle{Stability of Mutually Inclined Planetary Systems}
\shortauthors{Bhaskar et al.}
\begin{document}

\title{Dynamical and Secular Stability of Mutually Inclined Planetary Systems}

\author{Hareesh Gautham Bhaskar \altaffilmark{1}, Hagai Perets\altaffilmark{1}}
\affil{$^1$ Technion-Israel Institute of Technology}


\begin{abstract} 
Multiple analytical and empirical stability criteria have been derived in the literature for two planet systems. But, the dependence of the stability limit on the initial mutual inclination between the inner and outer orbits is not well modeled by previous stability criteria. Here, we derive a semi-analytical stability criteria for two planet systems, at arbitrary inclinations, in which the inner planet is a test particle. Using perturbation theory we calculate the characteristic fractional change in the semi-major axis of the inner binary $\beta=\delta a_1/a_1$ caused by perturbations from the companion. A stability criteria can be derived by setting a threshold on $\beta$ Focusing initially on circular orbits, we derive an analytical expression for $\beta$ for co-planar prograde and retrograde orbits. For non-coplanar configurations, we evaluate a semi-analytical expression. We then generalize to orbits with arbitrary eccentricities and account for the secular effects. Our analytical and semi-analytical results are in excellent agreement with direct N-body simulations. In addition, we show that contours of $\beta\sim0.01$ can serve as criteria for stability. More specifically, we show that (1) retrograde orbits are generally more stable than prograde ones; (2) systems with intermediate mutual inclination are less stable due to vZLK dynamics; and (3) mean-motion resonances (MMRs) can stabilize intermediate inclination secularly unstable regions in phase space, by quenching vZLK secular processes (4) MMRs can destabilize some of the dynamically stable regions. We also point out that these stability criteria can be used to constrain the orbital properties of observed systems and their age.
We also point out that these stability criteria can be used to constrain the orbital properties of observed systems (in particular inclination) and their age.	
\end{abstract}

\keywords{ Planetary systems}

\section{Introduction} 

The long-term stability of planetary systems is a well-studied problem. While it can be argued that no N $(>2)$-body system is stable for an arbitrarily long period \citep{arnol1964instability}, much progress can be made by using a definition of stability more suitable for astrophysical systems. Two definitions of stability are widely discussed in the literature: Hill stability and Lagrange stability (See \cite{daviesLongTermDynamicalEvolution2014} for a review). If the orbits of the planets are guaranteed to never cross, they are said to be Hill stable.  Meanwhile, if the orbits are protected against ejections or collisions, they are considered to be Lagrange stable.  It should be noted that Hill stability does not guarantee Lagrange stability, as secular perturbations can lead to ejections of planets even on non-crossing orbits. While proving Lagrange stability is much more difficult than proving Hill stability, it has been shown that in practice the Hill and the Lagrange stability limits lie close to each other (\cite{barnesStabilityLimitsExtrasolar2006}; \cite{barnesStabilityLimitsResonant2007}; \cite{deckRAPIDDYNAMICALCHAOS2012}).

% in the context of the restricted 3-body problem
Although various stability criteria had been derived for planetary systems using analytic tools (see below), the majority had considered co-planar systems, and provided criteria only for prograde or retrograde orbits. Very few studies explored more realistic arbitrary inclined systems, but could not provide correct results, as compared to full N-body simulation results (see e.g. Figs. 1 and 2). Moreover, \cite{grishinGeneralizedHillStabilityCriteria2017}  (see also \citep{Per+09}) have derived analytic stability criteria for star-planet-satellite systems, showing that secular dynamical evolution such as von Ziepel-Lidov-Kozai (vZLK) evolution \citep{von_zeipel_1910,lidov_evolution_1962,kozai_secular_1962} plays a key role in determining the stability of the systems for a wide range of intermediate mutual inclinations (typically 40$^\circ$-140$^\circ$, with dependence on initial eccentricity). This would suggest that such secular effects should be considered for the stability of other types of three-body systems. Here we use the analytic study of star - two-planet systems at arbitrary mutual inclinations to provide analytic stability criteria accounting both for the secular dynamical aspects, as well as mean-motion resonances which reproduces the empirical evidence from numerical few-body simulations.    
%Under certain approximations, simple analytical expressions can be derived for Hill stability criteria. Meanwhile, proving Lagrange stability is much more difficult, and no analytical criteria is known. Consequently, numerical integrations have been used to derive empirical stability criteria which are valid for tens of millions of orbits. Multiple studies have shown that in practice, the Hill and Lagrange stability limits lie close to each other (\cite{barnesStabilityLimitsExtrasolar2006}; \cite{barnesStabilityLimitsResonant2007}; \cite{deckRAPIDDYNAMICALCHAOS2012}). 
%An empirical criteria for Lagrange stability of unequal mass planets has been derived in \cite{kopparapuSTABILITYANALYSISSINGLEPLANET2010}
%\cite{petitHillStabilityAMD2018} rewrote the Hill stability criteria in terms of relative angular momentum deficit of the planets, and showed that planetary systems are Hill stable if their relative angular momentum is within a certain range.

Multiple studies have analyzed the Hill stability of star-two planet systems. In a circular restricted three-body system, an expression for Hill radius can be derived analytically by analyzing the zero velocity curves of the test particle (e.g., \cite{murraySolarSystemDynamics2000}). Planets inside the Hill radius are prevented from crossing it by the conservation of energy and angular momentum. The notion of Hill radius has been generalized \citep{marchalHillStabilityDistance1982} for arbitrary masses by using Sundman's inequality \citep{sundmanMemoireProblemeTrois1913}. \cite{gladmanDynamicsSystemsTwo1993} derived analytical Hill stability criteria for co-planar planetary systems using this approach. \cite{verasDynamicsTwoMassive2004} generalized the Gladman stability criteria to arbitrary mutual inclinations, however, this generalization provides reasonably correct results (in comparison with N-body calculations) only for very low-inclinations (see Fig. 2).    
%The Hill stability criteria for coplanar configurations has been refined to take into account the chaotic dynamics induced by the overlap of first order mean motion resonances \citep{deckFIRSTORDERRESONANCEOVERLAP2013}. 

Beyond compact systems, multiple empirical and semi-empirical stability criteria have been derived for hierarchical triple systems. Most of these works rely on an ensemble of N-body integrations. For instance, \cite{eggletonEmpiricalConditionStability1995} define a hierarchical triple system to be $n$-stable if none of the components of the triple escape the system or are exchanged in $10^{n}$ orbits of the outer companion. Using an ensemble of N-body simulations in which the initial conditions were sampled on a wide range of eccentricities (of both inner and outer binaries), mutual inclination, relative phase, and mass ratios, the authors derive empirical 2-stability criteria. \cite{mardlingTidalInteractionsStar2001} derive semi-empirical stability criteria by drawing an analogy between the triple stability and the chaotic energy exchange in the binary-tides problem. More recently, \cite{petrovichSTABILITYFATESHIERARCHICAL2015} analyzed the stability of hierarchical three-body systems using a large ensemble of long-term N-body integrations. They use the N-body results to train a Support Vector Machine classifier to search for an empirical stability boundary. For low mutual inclinations, this approach is shown to work better than the criteria previously derived by \cite{eggletonEmpiricalConditionStability1995}; \cite{gladmanDynamicsSystemsTwo1993}, and \cite{mardlingTidalInteractionsStar2001}; \cite{ginatAnalyticalStatisticalApproximate2021} provided an analytic statistical solution for the three-body chaotic problem (for non-extreme mass-ratios) and made use of another criterion. 

Multiple studies have analyzed the stability of planets in binary star systems (e.g.,
\cite{rablSatellitetypePlanetaryOrbits1988};
\cite{wiegertStabilityPlanetsAlpha1997};
\cite{holmanLongTermStabilityPlanets1999}; 
\cite{pilat-lohingerStabilityStypeOrbits2002};
\cite{pilat-lohingerStabilityLimitsDouble2003}; 
\cite{musielakStabilityPlanetaryOrbits2005};
\cite{haghighipourDynamicalStabilityHabitability2006}; 
\cite{quarlesLongTermStabilityTightly2018}; 
\cite{ballantyneLongtermStabilityPlanets2021}; \cite{decesareStabilityPlanetaryOrbits2021}). These planets can be classified into S-type (planets which orbit only one of the stars of the binary) and P-type (planets which orbit both stars of the binary). The setup we use in this paper is similar to the S-type configuration. \cite{wiegertStabilityPlanetsAlpha1997} studied the stability of P-type planets in the alpha centauri system using an ensemble of N-body simulations. In this study, the stability over the entire range of inclinations is studied (see their Figures 1-4). The authors find that the stability is a strong function of inclination, with retrograde orbits more stable than prograde orbits, and polar orbits being most unstable. In a newer study, \cite{quarlesLONGTERMSTABILITYPLANETS2016} confirm the results of \cite{wiegertStabilityPlanetsAlpha1997}, and find a similar dependence of the stability on the mutual inclination (see their Figure 1). They also point out the importance of MMRs in the stability of S-type planets. The general problem of the stability of S-type planets has also been studied in literature. For instance, \cite{rablSatellitetypePlanetaryOrbits1988} and \cite{holmanLongTermStabilityPlanets1999} use direct N-body integrations to deduce an empirical stability criteria for initially circular coplanar S-type planets. They derive a numerical fit for the critical semi-major axis of the planet beyond which it is unstable (as a function of binary eccentricity and mass ratio). More recently, \cite{quarlesOrbitalStabilityCircumstellar2020} improved the \cite{holmanLongTermStabilityPlanets1999} criteria using a larger ensemble of N-body simulations. A wide range of mass ratios, binary eccentricities and semi-major axis ratios were sampled in this study. The authors also changed the initial inclinations of the planetary orbits in their simulations. They also found that retrograde configurations are more stable than prograde orbits, with significant instability in the vZLK regime. Focusing on the vZLK regime, \cite{giupponeLidovKozaiStabilityRegions2017} found a similar result through their stability maps which showed the MEGNO values for a wide range of initial conditions (see their Figure 5). Beyond empirical criteria, the stability of P-type planets have also been studied using analytical and semi-analytical tools. For instance, \cite{quarlesLongtermStabilityPlanets2018} use secular perturbation theory to show that S-type planets that begin with eccentricity vectors near their forced values are generally stable. As pointed out by \cite{holmanLongTermStabilityPlanets1999}, their numerical fits cannot be directly applied to binary mass ratios less than 0.2. It should be noted that in this paper we focus on planetary regime with binary mass ratios of less than $10^{-3}$. Despite qualitative similarities, the results discussed above cannot be directly applied to planetary systems.

In the context of 3-body planetary systems (star + planet + planet), interactions between the planets can be generally classified into three categories. The planets can have close encounters, which can significantly change their eccentricities and semi-major axes. Sufficiently strong close encounters can destabilize, and trigger ejections from the system. When the planets are initialized within a few mutual hill radii of each other, they are more likely to have close encounters. At farther separations, perturbation theory can be used to study the dynamics of
the system. Depending on the configuration, both secular perturbations and Mean motion resonances (MMRs) can be important. MMRs are important when the orbital period ratio of the two interacting planets is commensurable. MMRs can be either stabilizing or destabilizing depending on the initial conditions. It should be noted that MMRs are important only within a certain distance (called resonance widths) of their nominal locations. Outside this range, secular perturbations dictate the long-term evolution of the system.

In the following, we first briefly overview the main stages of the calculations and the main results, as to provide a clear global picture. Readers interested in the detailed analysis can then follow our subsequent step-by-step detailed and technical analysis and derivation of the results.

%In this paper, we use perturbation theory to derive a semi-analytical stability criteria for two planet systems. More specifically, we analyze the oscillations in the semi-major axis of an inner low-mass planet caused by perturbations from a more massive outer planet. The planets are more likely to have close encounters if the amplitude of these oscillations is greater than a certain critical threshold. Our goal is to derive the dependence of the stability limit on the mutual inclination between the two planets, accounting for secular processes that can excite and change the planets' eccentricities and their mutual inclinations. We also study how MMRs can help stabilize the inner planet when the mutual inclination between the planets is large. In Section \ref{sec:setup} we describe the setup of the system. Our N-body results are discussed in Section \ref{sec:n-body}. We derive stability criteria for circular and eccentric orbits in Section \ref{sec:circstab} and \ref{sec:eccstab} respectively. The stabilizing effects of MMRs are described in Section \ref{sec:mmr}. We discuss our results and conclude in Section \ref{sec:conc}.

\section{overview}
In this paper, we use perturbation theory to derive semi-analytical stability criteria for two planet systems. More specifically, we analyze the oscillations in the semi-major axis of a low-mass inner planet caused by perturbations from a more massive outer planet. The planets are more likely to have close encounters if the amplitude of these oscillations is greater than a certain critical threshold. Our goal is to derive the dependence of the stability limit on the mutual inclination between the two
planets. We also study how MMRs can help stabilize the inner planet when the mutual inclination between the planets is large. 

We follow a step-by-step derivation, starting with simpler co-planar circular systems and exploring short timescale evolution, and then progressively relax the assumptions, to derive general stability criteria, first for non-co-planar systems, then for eccentric configurations, followed by the study of long-term secular effects, and finally also accounting for MMRs and their stabilizing effects due to quenching of secular evolution at high inclinations.

In Section \ref{sec:setup} we describe the setup of the system. Our setup is analogous to the circular restricted three-body problem. In practice, we find that our analytical and semi-analytical results are applicable as long as the mass ratio of the inner and the outer planets is less than $10^{-3}$ (see subsection \ref{subsec:mdep}). Our N-body results are discussed in Section \ref{sec:n-body}.  We find that the stability limit depends on the inclination of the system: retrograde orbits are more stable than prograde stable (see Fig \ref{fig:stabNbody}). At higher inclinations, the inner planets tend to be more unstable. This is due to eccentricity excitation caused by secular perturbations. We find that while the stability criteria previously derived in literature are {\emph not} consistent with N-body simulations, our semi-analytical criteria are in excellent agreement with them (see Fig. \ref{fig:compstablit}). To derive the stability criteria we calculate the characteristic fractional change in the semi-major axis of the inner planet ($\beta=\delta a_1/a_1$). In this calculation we use Laplace’s Equations (see Eqn. \ref{eqn:da1dt}), and assume that the orbital elements of the inner planet are constant. We initially focus on circular orbits (Section \ref{sec:circstab}). We find that an analytical expression for the maximum fractional change in the semi-major axis ($\beta_{circ}$) can be derived for co-planar orbits: $I_1=0$ and $180^\circ$ (see Eqn. \ref{eqn:dela1aneqn}). For arbitrary inclinations, we evaluate a semi-analytical expression. We find that both our analytical and semi-analytical results are in excellent agreement with N-body simulations on short timescales (see Figures  \ref{fig:compproan} and \ref{fig:comparbincan}). On longer secular timescales, the eccentricity of the inner binary can be excited, rendering our semi-analytical results inadequate. A stability criterion is derived by assuming that the inner orbit is stable when $\beta_{circ}<\beta_{crit}$. This circular stability criterion is consistent with N-body simulations on short timescales (see Figures \ref{fig:a1inc1an} and \ref{fig:compstabnbd1e2}). We generalize our derivation of $\beta$ to eccentric orbits ($\beta_{ecc}$) in Section \ref{sec:eccstab} (See \ref{fig:a1f1traj}). We find that $\beta_{ecc}$ has a strong dependence on the inclination and the argument of the pericenter of the inner binary (see Figure \ref{fig:a1f1traj}). For instance, for crossing orbits, $\beta_{ecc}$ can become unbound (see Fig. \ref{fig:dela1e1}). To define the stability limit that is valid for secular timescales, we calculate the maximum value of over secular timescales: $\beta_{sec}=max(\beta_{ecc})$ (see Fig. \ref{fig:secbetacalc}). We show that the stability limit defined by $\beta_{sec}>0.01$ is consistent with N-body simulations for secular timescales (See Fig. 
\ref{fig:nbodystabsecbeta}). In addition to secular effects, MMRs can also affect the stability of three body systems. Sometimes, the secular eccentricity excitation of the inner binary is suppressed by MMRs, which can stabilize the inner orbit (see Fig. \ref{fig:nbdyMMRHE}). The stabilizing effects of MMRs are described in Section \ref{sec:mmr} (see Fig. \ref{fig:contmmr21}). We discuss consistency of our results over different timescales (subsection \ref{subsec:ltsc}), comparison with observations (subsection \ref{subsec:compobs}), and the applicability of our results to systems with massive inner planets (subsection \ref{subsec:mdep}) in Section \ref{sec:discuss}. We conclude in Section \ref{sec:conc}.

\section{Setup}
\label{sec:setup}

In this paper, we focus on two-planet systems in which the mass of the inner planet ($m_1$) is much smaller than the mass of the outer planet ($m_2$). The mass of the host star is $m_0$. The orbital elements of the inner orbit are given by: $a_1$ (semi-major axis), $e_1$ (eccentricity), $I_1$ (inclination), $\omega_1$ (argument of pericenter), $\Omega_1$ (longitude of ascending node), $f_1$ (true anomaly), $M_1$ (mean anomaly) and $\lambda_1$ (longitude of pericenter). The orbital elements of the outer orbit are similarly defined, and are denoted by subscript 2. In our initial calculations, the inner planet is put on a near-circular orbit ($e_1=0.01$). The outer planet is assumed to be in a circular orbit ($e_2=0$). The reference plane is chosen such that the inclination and the longitude of the pericenter of the outer orbit are set to 0. These assumptions make our setup similar to the circular restricted three-body problem. The mean motion of the inner (outer) orbits is denoted by $n_1$ ($n_2$). In addition, the scale-invariant angular momentum of the inner orbit is given by $J=\sqrt{1-e_1^2}$, and its component along the orbit normal to the outer binary is $J_z = \sqrt{1-e_1^2} \cos{I_1}$.  It should be noted that since the outer planet is assumed to be on a circular orbit, by symmetry, $J_z=\sqrt{1-e_1}\cos{I_1}$ is a constant of motion.

In most of the calculations below, we set $a_1 = 1$ AU, $m_2=10^{-3} M_\odot$,  and $m_0$ = 1 $M_\odot$. It can be seen that the orbital period of the inner planet is $t_p = 1$ year. First order MMRs operate on timescales roughly given by $t_{res} \sim \sqrt{(m_0+m_1)/m_2} t_p \sim 32$ years. Meanwhile, secular perturbations operate on much longer timescales given by: $t_{sec} \sim (m_0/m_2)(a_2/a_1)^3 t_p \sim 3.4 \times 10^4 (a_2/1.5)^3$ years. 

\section{N-body Results}
\label{sec:n-body}

\begin{figure*}
	\centering
	\includegraphics[scale=0.7]{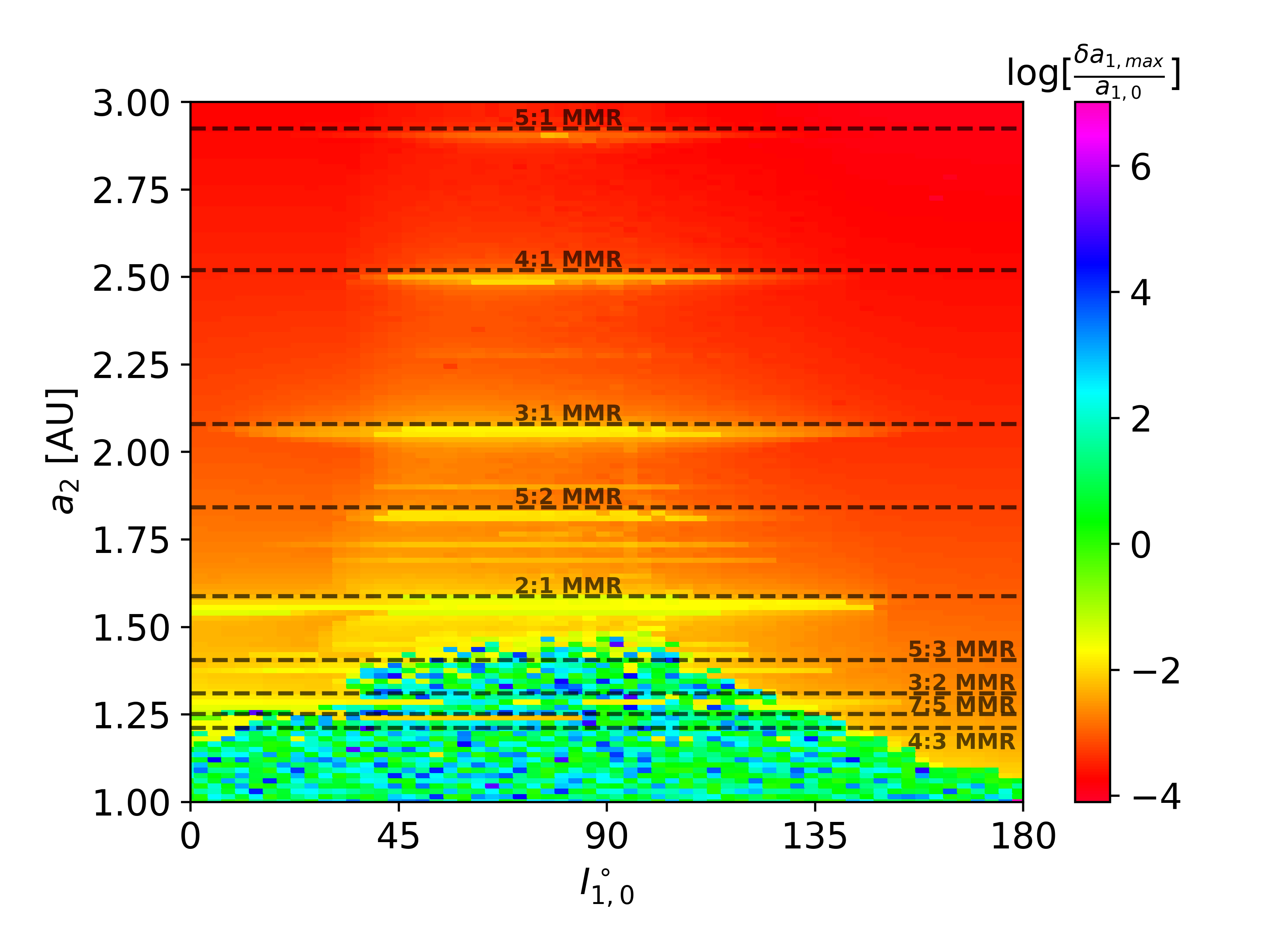}	
	\caption{The stability of two-planet systems as a function of mutual separation and inclination. The x-axis shows the semi-major axis of the outer planet ($a_2$), the y-axis shows the initial inclination of the inner planet ($I_{1,0}$), and the color shows the maximum fractional change in the semi-major axis of the inner planet after $10^6$ years of evolution. We can see that depending on the inclination of the inner orbit, the stability limit lies between 1.1 and 1.5 AU. Generally, retrograde orbits are more stable than prograde orbits. In addition, MMRs can significantly affect the stability of two-planet systems. The nominal locations of some of the MMRs are shown using the horizontal dashed lines. We can see that some of the MMRs can stabilize the inner orbits (see the region near 7:5 MMRs). We choose the following initial condition to make this plot: $a_1 = 1$ AU, $e_1=0.01$, $e_2=0.0$, $m_0 = 1 M_\odot, m_1=10^{-6} M_\odot$ and $m_2=10^{-3} M_\odot$. Other orbital elements including $\omega_1,\Omega_1, f_1$ and $f_2$ are chosen uniformly between 0 and $2\pi$.}
	\label{fig:stabNbody}
\end{figure*}

In this section, we will discuss the main results from our ensemble of N-body simulations. For our simulations, we use the IAS15 integrator from rebound \citep{reinREBOUNDOpensourceMultipurpose2012}. %HBP: Add a reference to rebound paper
We choose a time-step of 1/20th the orbit period of the inner binary. We stop our integration when the inner planet either becomes unbound ($a_1 < 0$) or ceases to be on an elliptical orbit ($e_1 \geq 1$). If neither of these conditions is satisfied in $t_{max} \in (100, 10^8)$ years, we end our integration. 

Figure \ref{fig:stabNbody} shows results from our N-body simulations. The semi-major axis of the outer planet is shown on the y-axis, and the initial inclination of the inner planet is shown on the x-axis. The color shows the maximum fractional change in the semi-major axis of the inner planet after $t_{max}=10^6$ years of evolution. We can see that near co-planar prograde orbits  ($I_1 < 20^\circ$) are unstable ($\delta a_1/a_1 \sim 1$) when $a_2 < 1.3$ AU. At higher inclinations ($20^\circ < I_1 <160^\circ$), secular perturbations excite the inner eccentricity triggering instability even when the perturber is farther away ($a_2 \sim 1.5$ AU). Also, the inner binary tends to be more stable on retrograde orbits. For instance, we can see that when $I_1 \sim 180^\circ$ the inner binary becomes unstable only when the perturber is at $a_2 = 1.1$ AU.

MMRs can also affect the stability of two-planet systems. The nominal location of some of the MMRs is shown in Figure \ref{fig:stabNbody} as dashed black lines. Figure \ref{fig:stabNbody} shows that most of the misaligned planetary systems with $a_2<1.5$ AU are unstable. An exception to this general trend is planets initialized near the nominal location of MMRs. For instance, we can see that some of the inner planets initialized at the nominal location of 7:5 MMRs are stable. As we will see later, the stability of an inner orbit initialized near MMRs depends on the initial phase of their resonance angles.

\begin{figure}
	\centering
	\includegraphics[scale=0.5]{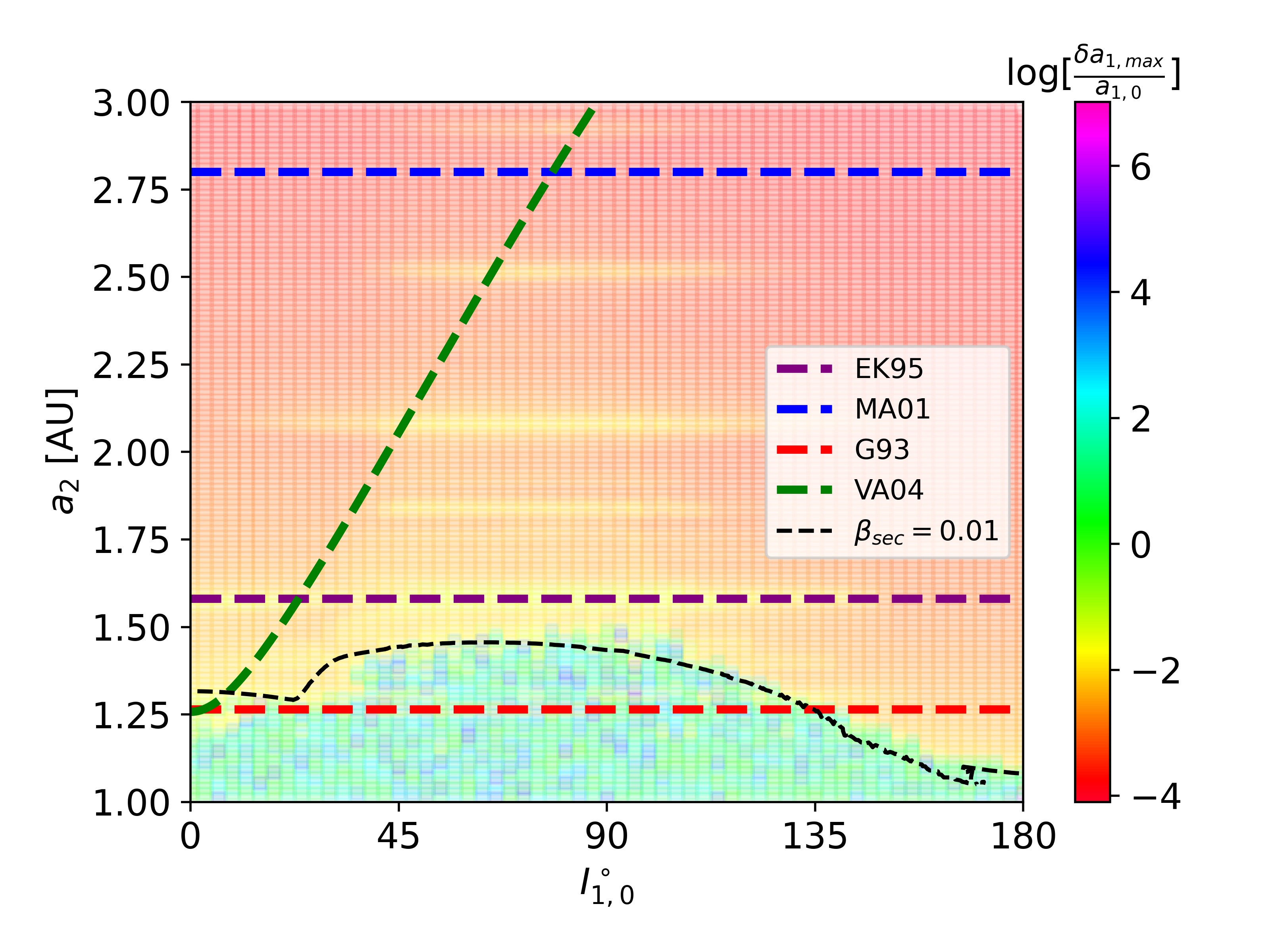}
	\caption{Comparison of our stability criteria to previous criteria from literature and with N-body simulations. This figure is similar to Figure \ref{fig:stabNbody}, except the dashed lines show the stability criteria previously derived in literature. We can see that none of the stability criteria shown here are consistent with the dependence of the stability limit on the initial mutual inclination seen in N-body simulations. In contrast, our derived criterion, shown in black dotted line, well captures the stability region.
    }
	\label{fig:compstablit}
\end{figure}

Figure \ref{fig:compstablit} compares the stability criteria previously derived in literature with our N-body simulations. The stability criteria derived in \cite{gladmanDynamicsSystemsTwo1993a} is shown in the figure as the red dashed line. We can see that it agrees with N-body simulations only for low initial mutual inclinations ($I_1<20^\circ$). This is consistent with the fact that the Gladman criteria were originally derived for co-planar prograde planetary systems. \cite{verasDynamicsTwoMassive2004} generalized the Gladman criteria to arbitrary mutual inclinations. It is shown in the figure as the green dashed line. We can see that it too cannot reproduce the features seen in our N-body results; while the stability criteria is consistent with N-body simulations for low inclination prograde orbits, it fails at higher inclinations and for retrograde orbits. Also, since it is based on the Hill stability criteria, it cannot account for MMRs. The empirical stability criteria derived in \cite{eggletonEmpiricalConditionStability1995a} and \cite{mardlingTidalInteractionsStar2001e} are shown in the figure as the purple and the blue dashed lines respectively. These criteria were derived for comparable mass ratio hierarchical triple systems, and hence a direct comparison with our N-body simulations may not be appropriate. Nevertheless, we can see that these criteria do not have a strong dependence on the initial mutual inclination (which would be expected to be important for near-equal mass cases too), and hence do not capture the asymmetry in the stability limit seen in our N-body simulations. 

In contrast, our analytic stability criteria, derived below, well reproduce the N-body results.  

\section{Derivation of Stability Criteria for circular orbits}
\label{sec:circstab}
The rest of the paper is devoted to deriving semi-analytical stability criteria using perturbation theory. We first focus on circular and co-planar orbits, and then generalize to orbits with arbitrary inclinations and eccentricities.

\subsection{Co-planar orbits}
 The rate of change of the semi-major axis of the inner orbit due to perturbations from the outer companion is given by Lagrange’s Equation:
\begin{equation}
	\label{eqn:da1dt}
	\frac{d a_1}{dt} = \frac{2}{n_1a_1} \frac{\partial R}{\partial M_1}
\end{equation}

where $R$ is the disturbing function \citep{murraySolarSystemDynamics2000}. The total disturbing function can be written as the sum of the direct ($R_d$) and the indirect ($R_I$) parts: $R=R_d+R_I$. The expression for $R_d$ and $R_I$ are given by, 
\begin{eqnarray}\label{eqn:distfn}
	R_d &=& G m_2\frac{a_2}{|\vec{r}_2- \vec{r}_1|} \\
	R_I &=& -G m_2\frac{|\vec{r}_2|}{a_2} \left(\frac{a_1}{|\vec{r}_1|}\right)^2 \cos{\psi}
\end{eqnarray}

where $\vec{r}_1$  and $\vec{r}_2$  are the relative position vectors of the inner planet and the companion with respect to the host star. $\psi$ is the angle between the two position vectors (i.e., $\cos{\psi} = \vec{r}_1.\vec{r}_2/(|\vec{r}_1||\vec{r}_2|)$), and $G$ is the universal constant of gravity. The expression for $\cos {\psi}$ is given by: 

\begin{eqnarray}\label{eqn:psieqn}
	\cos{\psi} &=& \frac{1}{2} ( (1 + \cos{I}) \cos(f_1 - f_2 + \omega_1 + \Omega_1) \nonumber \\
	&&  -((-1 + \cos{I}) \cos(f_1 + f_2 + \omega_1 - \Omega_1)))
\end{eqnarray}

For circular orbits, $r_1=a_1$, $r_2=a_2$, $f_1  = n_{1}t + f_{10} $ and $f_2 = n_{2}t + f_{20}$. Also, $|\vec{r}_2- \vec{r}_1| = \sqrt{a_1^2+a_2^2-2a_1 a_2\cos{\psi}}$. Here, $f_{10}$ and $f_{20}$ are the initial true anomalies of the inner and the outer orbits. Putting it in Lagrange’s equation, we get:
\begin{equation}\label{eqn:lagda1dt}
	\frac{d a_1}{dt} = \frac{G m_2}{a_2^2 n_1} \left( -1 + \frac{1}{(1+ \alpha^2 -2\alpha\cos{\psi})^{3/2}}\right) \frac{d \cos{\psi}}{d f_1},
\end{equation}
where $\alpha = a_1/a_2$. In this section, we are only interested in the evolution of the semi-major axis of the inner planet on timescales comparable to its orbital period. Hence, while evaluating the right-hand side of Eqn. \ref{eqn:lagda1dt}, we assume that the orbital elements of the inner orbit are constant. Consequently, $\alpha$ is also assumed to be a constant. This is a good approximation as long as the changes in the semi-major axis of the inner planet are small ($\delta a_1/a_1 \lesssim 0.01$). 

From Eqn. \ref{eqn:lagda1dt}, we can see that the semi-major axis of the inner binary is maximized when $\frac{d a_1}{dt} =0 \Rightarrow \alpha = 2\cos{\psi}$. Using this, an analytical expression for the maximum value of $\delta a_1 (= a_1(t)-a_1 (t=0))$ can be derived for co-planar orbits i.e., $I =0^\circ$ and $180^\circ$:
\begin{eqnarray}\label{eqn:dela1aneqn}
	\beta_{circ,cr/cp} &=& \frac{\text{Max}[\delta a_{1_\pm}]}{a_{1}} \nonumber \\
                     &=& \frac{G m_2}{a_2^2 a_1 n_1 (n_1 \pm n_2)\alpha }  \left(3- \Delta_{\pm}^2 -\frac{2}{\Delta_{\pm}} \right)
\end{eqnarray}
where $\Delta_{\pm} = \sqrt{ 1+\alpha^2-2\alpha \cos(f_{10} \pm f_{20} + \omega_1 +\Omega_1)}$. Here, $\beta_{circ,cr} (\beta_{circ,cp})$ is the fractional change in the semi-major axis of the inner orbit for $I_1=180^\circ$ ($I_1=0^\circ$). 

A stability criteria can be derived by setting a threshold on the {\it characteristic} fractional change in the semi-major axis of the inner binary i.e., the binary is assumed to be stable if:
\begin{equation}\label{eqn:stabcrit}
	\beta=\frac{\delta a_1}{a_1} < \beta_{crit}
\end{equation}
where $\beta_{crit}$ is generally taken to be of the order $10^{-2}$. We can derive an analytical stability limit for co-planar orbits by setting $\beta=\beta_{circ,cp/cr}$ in the above equation. From Eqn. \ref{eqn:dela1aneqn}, we can see that retrograde orbits are more stable as compared to prograde orbits. 

\begin{figure}
	\centering
	\includegraphics[scale=0.5]{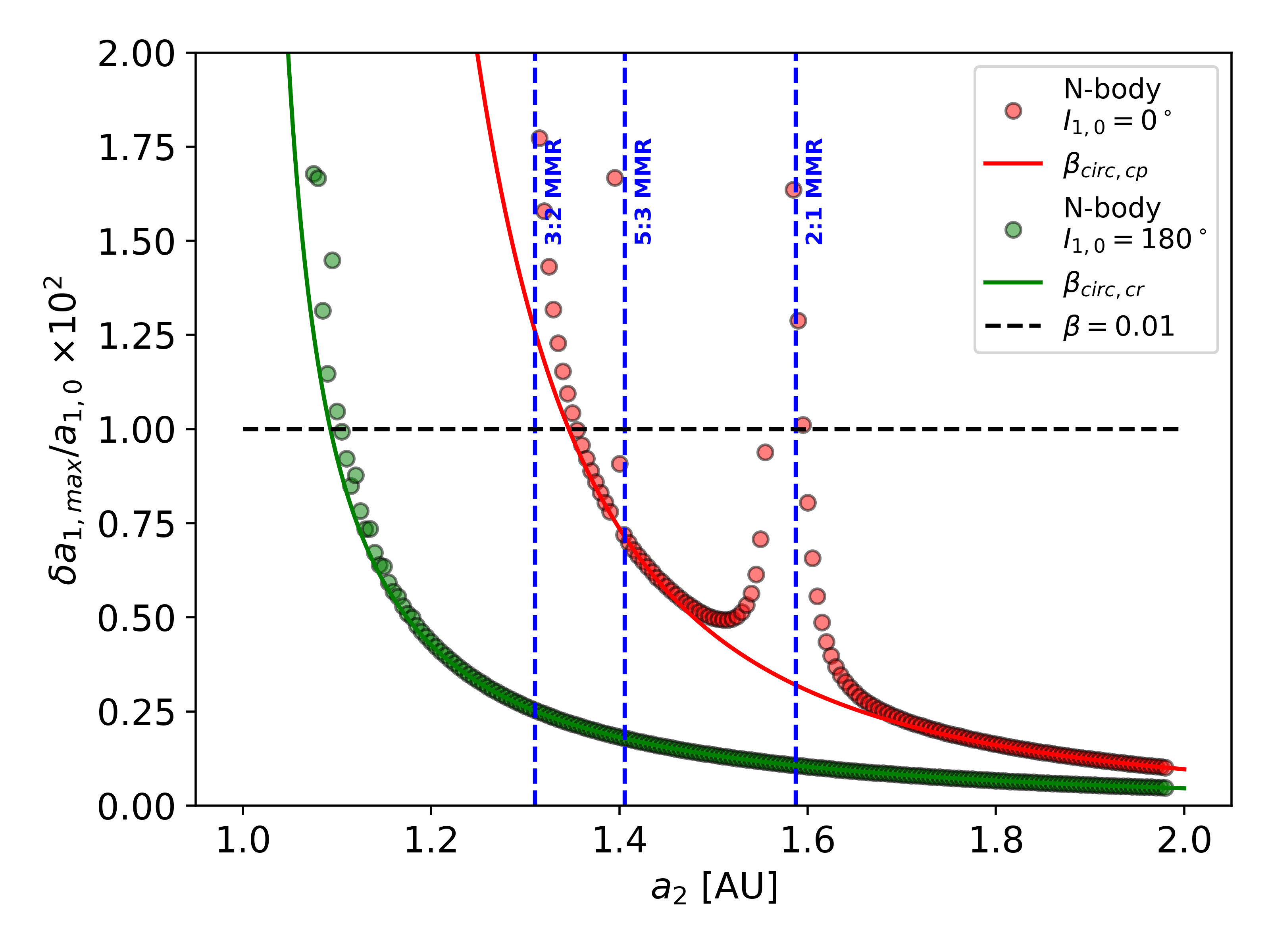}
	\caption{ The fractional change in the semi-major axis of the inner orbit ($\delta a_1/a_1$) as a function of the semi-major axis of the companion ($a_2$). The inner orbit is initialized on a co-planar orbit with an inclination ($I_1$) of $0$ (shown in red) and $180^\circ$ (shown in green). The N-body results are shown using dots, the analytical results are shown using solid lines, and the location of MMRs is shown using vertical dashed lines. We can see that the analytical expression agrees with N-body simulations except near the MMRs. A stability limit can be defined by setting a threshold on the fractional change in the semi-major axis ($\beta < \beta_{crit}$). The stability criteria given by $\beta_{crit} =10^{-2}$ is shown in the figure as the dashed black line.  We choose the following initial condition to make this plot: $a_1 = 1$ AU, $e_1=0.0$, $e_2=0$, $m_0 = 1 M_\odot, m_1=0$, and $m_2=10^{-3} M_\odot$. Other orbital elements including $\omega_1,\Omega_1, f_1$ and $f_2$ are chosen uniformly between 0 and $2\pi$. }
	\label{fig:compproan}
\end{figure}

%\begin{figure}
	%\centering
	%\includegraphics[scale=0.5]{figures/compRetroAnalytical.png}
	%\caption{Same as Figure 10, except for retrograde test particle orbit with inclination of 180 degrees.  We can again see that %beyond the stability limit, N-body simulations agree with the analytical criteria. The threshold is now breached at 1.1 AU, closer %to the test particle as compared to the prograde orbits.}
	%\label{fig:compretan}
%\end{figure}

Figure \ref{fig:compproan} compares Eqn. \ref{eqn:dela1aneqn} with N-body simulations. The dots show results from N-body simulations, and the solid lines show results from Eqn. \ref{eqn:dela1aneqn} for prograde and retrograde orbits in red and green respectively. The stability threshold of $\beta_{crit} = 0.01$ is shown using the black dashed line. The nominal locations of various MMRs are shown using the blue dashed lines. At larger values of $a_2$ ($ \gtrsim  1.4$ AU for $I_1=0$, and $ \gtrsim  1.2$ AU for $I_1=180^\circ$ ), we can see that the analytical results are in agreement with N-body simulations except near MMRs. The disagreement near MMRs is because in our derivation of Eqn. \ref{eqn:dela1aneqn} we assume that the eccentricity and the inclination of the inner orbit is exactly zero. This removes all the resonant terms in our derivation. Meanwhile, the eccentricity of the inner orbit cannot be set to zero at all times in N-body simulations. We can see that the stability threshold is reached for co-planar prograde orbits at 1.35 AU (except near 2:1 and 5:3 MMRs at 1.58 and 1.4 AU respectively), and for co-planar retrograde orbits at 1.1 AU, consistent with the N-body simulations (see Fig. \ref{fig:stabNbody})

\subsection{Arbitrary mutual inclinations}
\begin{figure}
	\centering
	\includegraphics[scale=0.5]{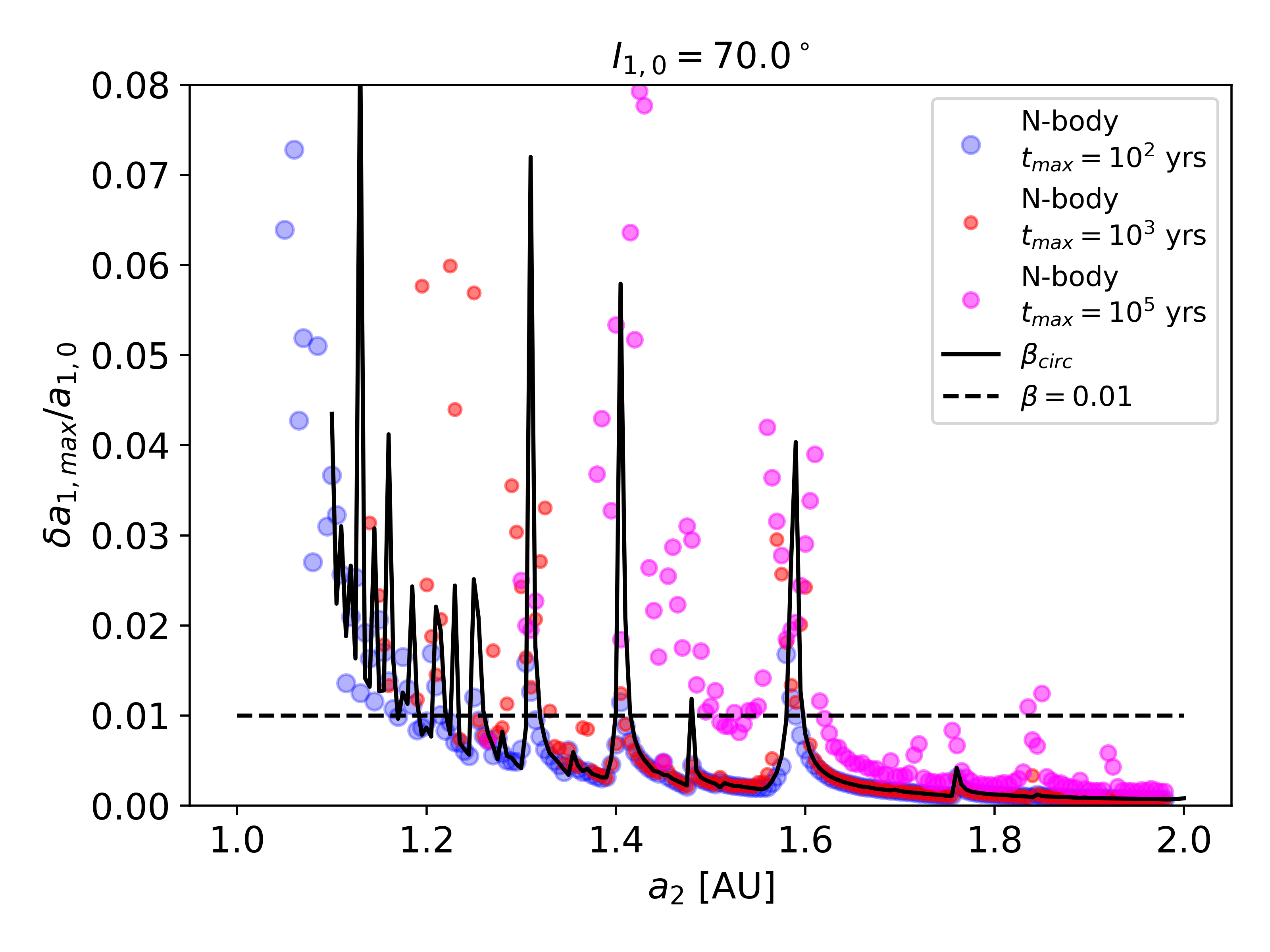}
	\caption{
The fractional change in the semi-major axis of the inner orbit ($\delta a_1/a_1$) as a function of the semi-major axis of the companion ($a_2$). The inner orbit is initialized at an inclination of $I_1=70^\circ$. The semi-analytical results are shown using the solid black line, and the N-body results are shown using dots.  The maximum fractional change in the semi-major axis of the inner orbit after 100, $10^3$, and $10^5$ years of evolution are shown using the blue, red, and pink dots respectively. On short timescales ($<10^3$ years) can see that the semi-analytical and N-body results agree with each other. On longer timescales, secular perturbations can significantly change the orbital elements of the inner orbit. Consequently, the semi-analytical approach used in this section, not yet including secular effects, is inadequate for the longer timescales.  We choose the following initial condition to make this plot: $a_1 = 1$ AU, $e_1=0.0$, $I_1=70^\circ$, $e_2=0$, $m_0 = 1 M_\odot, m_1=0$ and $m_2=10^{-3} M_\odot$. Other orbital elements including $\omega_1,\Omega_1, f_1$ and $f_2$ are chosen uniformly between 0 and $2\pi$.}
	\label{fig:comparbincan}
\end{figure}

We now generalize the above analysis to arbitrary inclinations. While a simple analytical expression is not possible, we numerically calculate the maximum change in the semi-major axis of the inner orbit ($\delta a_{1,max}$) using Eqn. \ref{eqn:lagda1dt}. The maximum {\it fractional} change in the semi-major axis of the inner orbit is then given by $\beta_{circ} = \delta a_{1,max}/a_1$. The solid black line in Figure \ref{fig:comparbincan} shows $\beta_{circ}$ as a function of the semi-major axis of the companion.  Results from N-body simulations after $10^2$, $10^3$, and after $10^5$ years of evolution are shown using the blue, red, and pink dots respectively. As can be seen, the semi-analytical results agree with N-body simulations on short timescales (red and blue dots). Meanwhile, on longer timescales, the change in the semi-major axis of the inner orbit is higher in the N-body simulations as compared to our analytical results. This is because secular effects, which become important on timescales longer than $10^4$ years, can cause large eccentricity and inclination variations in the orbits of misaligned inner orbits. Our semi-analytical derivation, which assumes that the inner orbit is on a constant circular orbit, is hence inadequate. It should be noted that for near co-planar orbits, secular effects do not significantly change either the eccentricity or the inclination of the inner orbit. Consequently, our semi-analytical results for $I_1 \lesssim 40^\circ$ and $I_1 \gtrsim 140^\circ$ are consistent with N-body simulations on longer timescales.

\begin{figure}
	\centering
	\includegraphics[scale=0.5]{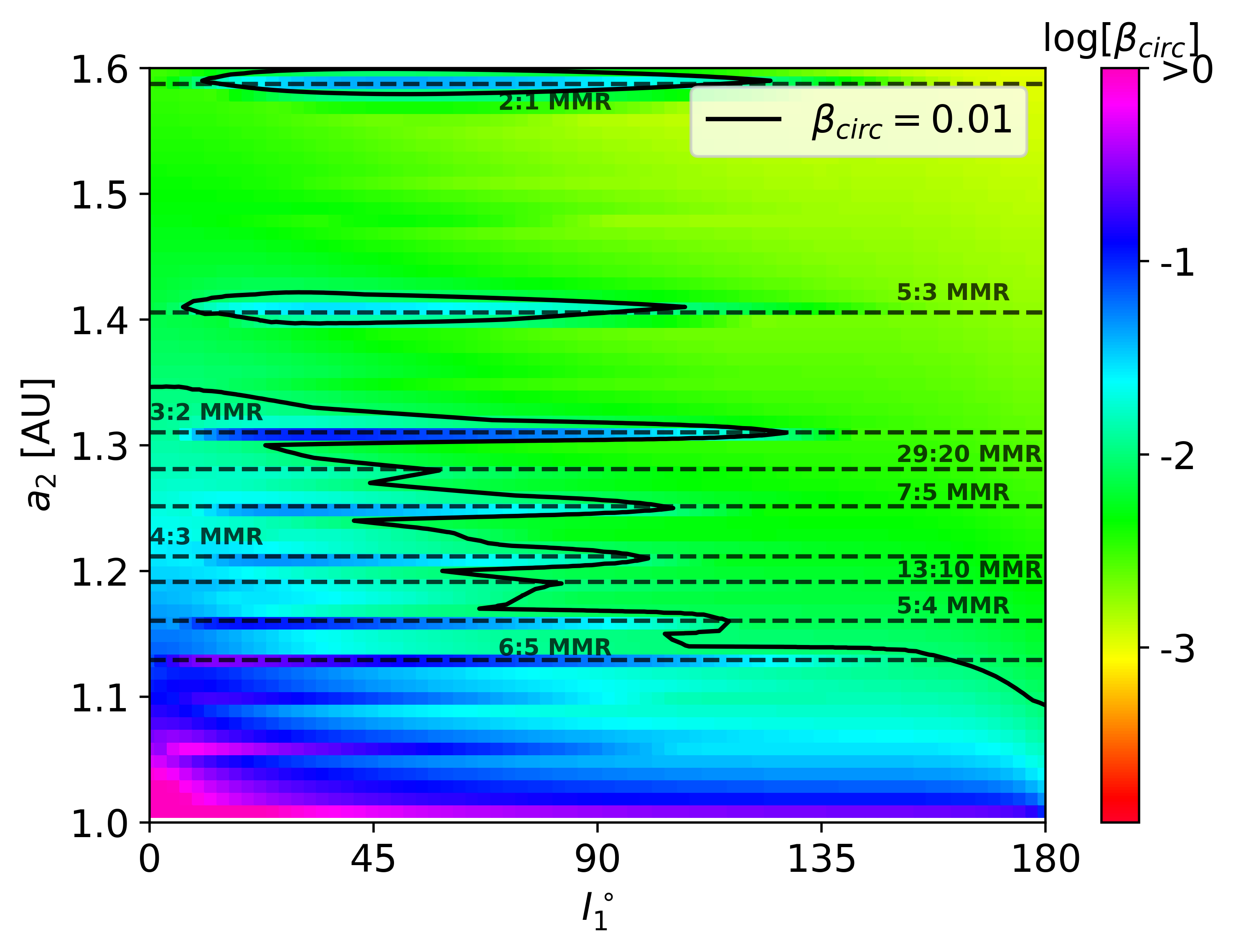}
	\caption{
The semi-analytical fractional change in the semi-major axis of the inner orbit ($\beta_{circ}$, from Eqn. \ref{eqn:lagda1dt}) as a function of the semi-major axis of the companion ($a_2$), and the inclination of the inner orbit ($I_1$).  The contour of $\beta=0.01$ is shown using the black line. The irregular shape of the contour is due to MMRs. We can see that $\beta=0.01$ is satisfied when $a_2=1.35$ AU for $I_1=0$, and $a_2=1.1$ AU for $I_1=180^\circ$. This is similar to the stability limit seen in our N-body results (see Fig. \ref{fig:stabNbody}). We chose the following initial conditions to make this plot: $a_1 = 1$ AU, $e_1=0$, $\Omega_1=0$, $e_2=0$, $f_1=f_2=0$, $m_0 = 1 M_\odot, m_1=0$, and $m_2=10^{-3} M_\odot$. }
	\label{fig:a1inc1an}
\end{figure}

Figure \ref{fig:a1inc1an} shows $\beta_{circ}$ (calculated using Eqn. \ref{eqn:lagda1dt}) as a function of both the semi-major axis of the companion (y-axis) and the mutual inclination (x-axis).  As expected, $\beta_{circ}$ is higher when the companion is close to the inner orbit. We can see that in general, retrograde orbits are more stable than prograde orbits. We can also see regions of instability embedded inside generally stable regions. These regions correspond to the location of MMRs. The stability criteria are given by the Eqn. \ref{eqn:stabcrit} can be obtained by setting $\beta = \beta_{circ}$. The contour corresponding to $\beta_{circ}=10^{-2}$ is shown in the figure as the solid black line. The irregular shape of the contour is due to the presence of multiple MMRs near the stability limit ($a_2<1.4$ AU).

\begin{figure*}
	\centering
	\includegraphics[scale=1.0]{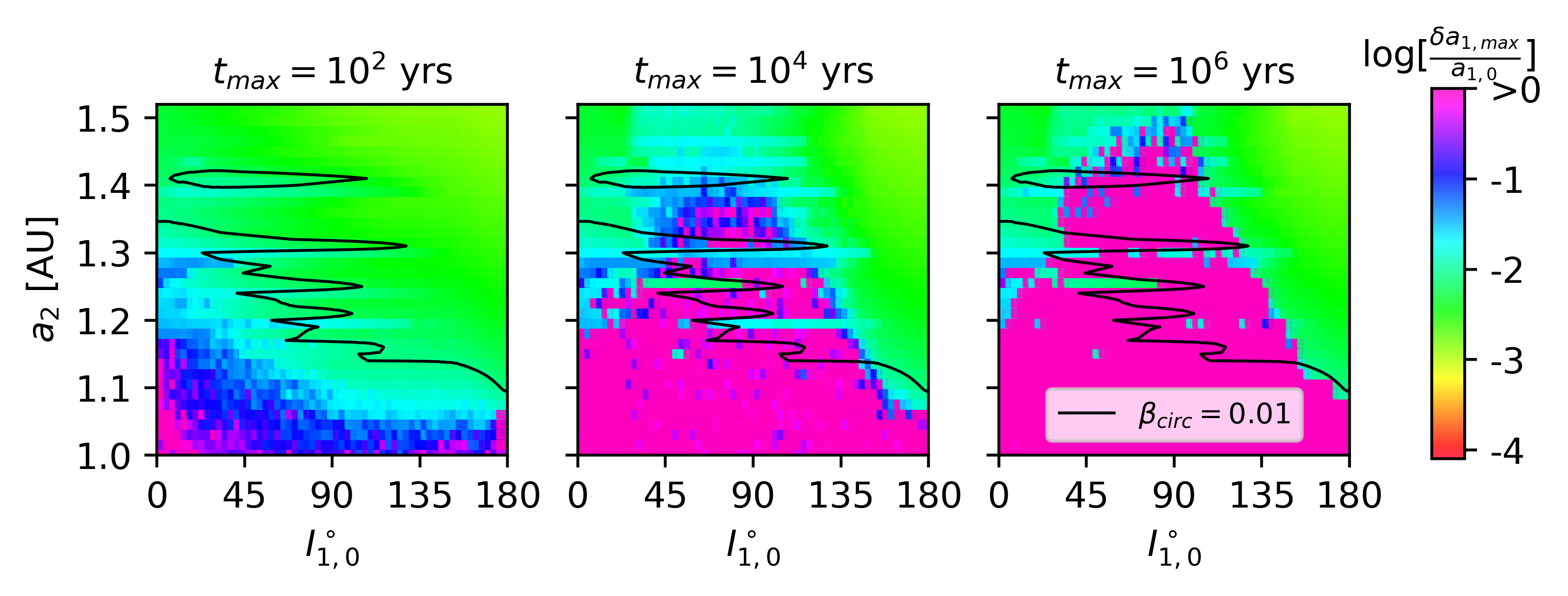}
	\caption{
 Comparison of the semi-analytical stability criteria $\beta_{circ}=0.01$ with direct N-body simulations. The color shows the fractional change in the semi-major axis of the test particle as calculated by N-body simulations after 100 (left panel), $10^4$ (middle panel), and $10^6$ (right panel) years of evolution. The black lines show the contour of $\beta_{circ}=0.01$ from Figure \ref{fig:a1inc1an}. We can see that the contour is consistent with N-body simulations in the left panel. On longer timescales (middle and right panels), the eccentricity of misaligned inner orbits ($I_1>40^\circ$ and $I_1<140^\circ$) is excited due to secular perturbations, and the semi-analytical results derived in this section are inadequate. Nevertheless, the semi-analytical results are consistent with N-body results for near more co-planar orbits ($I_1<40^\circ$ and $I_1>140^\circ$), at which secular effects play a little role in the evolution. We chose the following initial conditions for N-body simulations: $a_1 = 1$ AU, $e_1=0.01$, $e_2=0$, $m_0 = 1 M_\odot, m_1=10^{-6} M_\odot$ and $m_2=10^{-3} M_\odot$. Other angles including $\omega_1,\Omega_1, f_1$ and $f_2$ are chosen uniformly between 0 and $2\pi$.
 }
	\label{fig:compstabnbd1e2}
\end{figure*}

%\begin{figure}
%	\centering
%	\includegraphics[scale=0.5]{figures/compStabAnNbdy1e4.png}
%	\caption{Same as Figure \ref{fig:compstabnbd1e2}, but the color now shows the fractional change in the semi-major axis of the test particle after $10^4$ years. Secular effects increase the eccentricity for initially misaligned orbit ($20<I_{mut}<160$). In this regime, the semi-analytical criteria is not accurate, since the derivation of the criteria assumes a circular test particle orbit.}
%	\label{fig:compstabnbd1e4}
%\end{figure}

%\begin{figure}
%	\centering
%	\includegraphics[scale=0.5]{figures/compStabAnNbdy1e6.png}
%	\caption{Same as Figure \ref{fig:compstabnbd1e4}, but the color now shows the fractional change in the semi-major axis of the test particle after $10^6$ years.}
%	\label{fig:compstabnbd1e6}
%\end{figure}

We now compare our semi-analytical results with N-body simulations. Figure \ref{fig:compstabnbd1e2} is similar to Figure \ref{fig:a1inc1an}, except it shows results from N-body simulations: the color shows the fraction change in the semi-major axis from N-body integration after 100 (left panel), $10^4$ (middle panel) and $10^6$ (right panel) years of evolution. The contour from Figure \ref{fig:a1inc1an} is shown using solid black lines. We can see that the semi-analytical contours agree well with N-body simulations run for $t_{max}=100$ years (left panel). On longer timescales (middle and right panels) secular effects become important for misaligned orbits ($20 ^\circ<I_{1}<160^\circ$). Secular perturbations excite the eccentricity to large values, which renders the semi-analytical approach discussed in this section inadequate. Outside this inclination range, the eccentricity remains low and the contours predict the stability limit well.

%\begin{figure}
%	\centering
%	\includegraphics[scale=0.5]{figures/a1_f1_traj.png}
%	\caption{The change in the semi-major axis of the test particle as a function of the true anomaly. The red line shows results from N-body simulations and the blue line shows results of semi-analytical calculation. We can see that during the apocenter passage there is a jump in the semi-major axis of the test particle. }
%	\label{fig:a1f1traj}
%\end{figure}

%\begin{figure}
%	\centering
%	\includegraphics[width=1.0\linewidth,height=0.75\linewidth]{figures/dela1f2relation.png}
%	\caption{The change in the semi-major axis of the test particle as a function of initial true anomaly of the companion. The semi-analytical results (from Eqn. \ref{eqn:dela1ecc}) are shown in orange and the N-body results are shown in blue. We can see that they agree with each other. The mean value of the semi-major axis kick is shown using the horizontal red an green lines. }
%	\label{fig:dela1f2}
%\end{figure}

\begin{figure}
	\centering
	\includegraphics[scale=0.5]{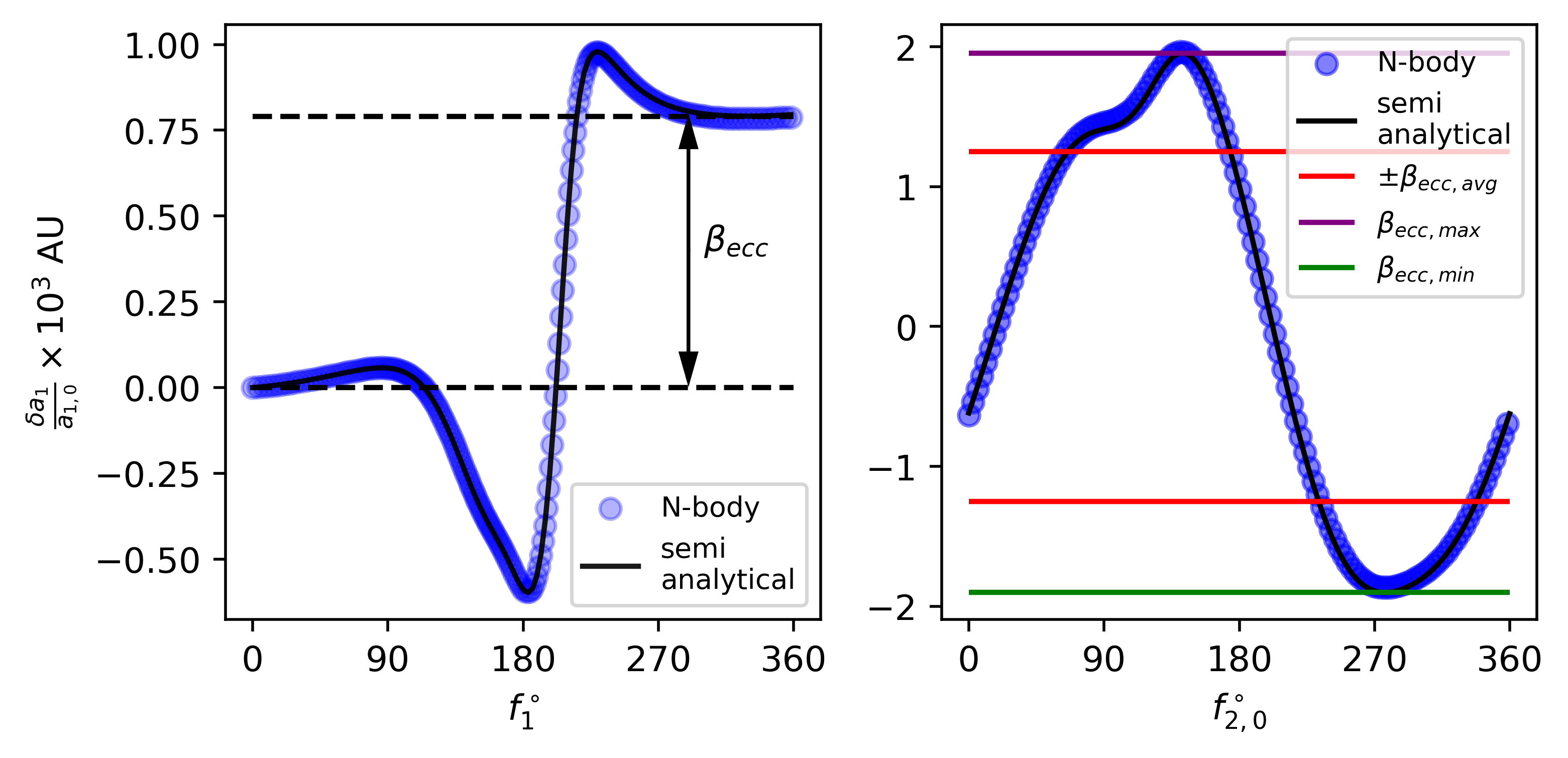}
	\caption{ The left panel shows the fractional change in the semi-major axis of the inner orbit as a function of the true anomaly of the inner orbit. The blue dots show results from N-body simulations, and the black line shows the semi-analytical calculation (Eqn. 9). We can see that during the apocenter passage, there is a kick in the semi-major axis of the inner orbit. The right panel shows the change in the semi-major axis of the test particle as a function of the true anomaly of the companion during pericenter passage ($f_{2,0} = f_2 (f_1 = 0)$). The semi-analytical (from Eqn. 9), and N-body results are again shown using the black line and the blue dots respectively, showing excellent agreement. The average value of the semi-major axis kick is shown using the horizontal red lines. The maximum positive and negative kicks are shown using the purple and green lines. We choose the following initial conditions for N-body simulations: $a_1 = 1 $ AU, $e_1=0.8$, $I_1=80^\circ$, $\Omega_1=5.89$, $\omega_1=5.18$, $a_2=1.5$ AU, $e_2=0$, $m_0 = 1 M_\odot, m_1=0$ and $m_2=10^{-3} M_\odot$. In the left panel, $f_{2,0} = 0$.
    }
	\label{fig:a1f1traj}
\end{figure}

\section{Stability criteria for eccentric orbits}
\label{sec:eccstab}

We now extend the above analysis to eccentric orbits. For eccentric orbits, it is more useful to use the derivative of the semi-major axis with respect to the true anomaly. The equation is given by:
\begin{equation}\label{eqn:da1df1}
	\frac{d a_1}{d f_1} = \frac{2}{n^2_1a_1} \frac{\partial R (f_1,f_2,\omega_1,\Omega_1,I_1)}{\partial f_1} \bigg \rvert_{f_2}
\end{equation}
It should be noted that the true anomaly of the companion ($f_2$) depends on the true anomaly of the test particle ($f_1$) through Euler's equation: $f_2 (f_1)=f_{20} + n_2t=f_{20} + (n_2/n_1)(E_1 -e_1\sin{E_1})$, where $E_1$ is eccentric anomaly of the inner orbit, and $\sin(E_1) = \frac{\sqrt{1-e_1^2}\sin{f_1}}{1+e_1\cos{f_1}}$. The change in the semi-major axis over one orbital period is given by,
\begin{equation}\label{eqn:dela1ecc}
 \beta_{ecc} = \frac{\delta a_1}{a_1} = \frac{2}{n^2_1a_1} \int_0 ^{2\pi} \frac{\partial R (f_1,f_2,\omega_1,\Omega_1,I_1)}{\partial f_1} \bigg \rvert_{f_2} df_1 
\end{equation}

It should be noted that unlike in the circular orbit case \footnote{see the definition of $\beta_{circ}$ in section \ref{sec:circstab}}, $\beta_{ecc}$ is not the maximum fractional change in the semi-major axis of the inner orbit. We use the above expression of $\beta$ as it is more physically intuitive, and also easier to calculate. The left panel of Figure \ref{fig:a1f1traj} shows the fractional change in the semi-major axis of the inner orbit as a function of the true anomaly of the inner orbit. The results from Eqn. \ref{eqn:dela1ecc} are shown using the black line. N-body results are shown as blue dots. We can see that both results agree with each other. We can see a jump in the semi-major axis during the apocenter passage. This is because the inner particle when on an eccentric orbit spends most of the time at the apocenter. 

From Eqns. \ref{eqn:da1df1} and \ref{eqn:dela1ecc} we can see that the magnitude of the semi-major axis kick also depends on the true anomaly of the companion during the pericenter passage of the test particle ($f_{2,0} = f_2(f_1=0)$). The right panel of Figure \ref{fig:a1f1traj} shows the change in the semi-axis of the test particle as a function of $f_{2,0}$. Results from Eqn. \ref{eqn:dela1ecc} and N-body simulations are shown using the black line and the blue dots respectively. We can see that they agree with each other. The curve for $\delta a_1$ is periodic in $f_{2,0}$. Since the perturber is on a circular orbit, $f_{2,0}$ is uniformly distributed between 0 and $2\pi$. The magnitude of average kick ($<\delta a_1>$) can be calculated numerically by integrating the absolute value of $\beta_{ecc}$ (from Eqn. \ref{eqn:dela1ecc}) over $f_{2,0}$ and dividing by $2\pi$. Figure \ref{fig:a1f1traj} shows the average kick magnitude using the horizontal red lines. The maximum positive and negative kicks are shown using the purple and green lines.

%\begin{figure}
%	\centering
%	\includegraphics[width=1.0\linewidth,height=0.75\linewidth]{figures/comp_dela1_ecc_inc80.png}
%	\caption{Comparison of semi-analytical expression with N-body simulations for a wide range of test particle eccentricities. The y-axis shows the change in the semi-major axis of the test particle and the x axis shows the eccentricity of the test particle. Results from N-body simulations are shown using the blue dots. The N-body integrations are run for 1 orbital period. The maximum, minimum abd average kick envelopes are shown using the blue, orange and red lines. They are derived semi-analytically using Eqn. \ref{eqn:dela1ecc}.}
%	\label{fig:dela1e1}
%\end{figure}

\begin{figure*}
	\centering
	\includegraphics[width=1.0\linewidth,height=0.38\linewidth]{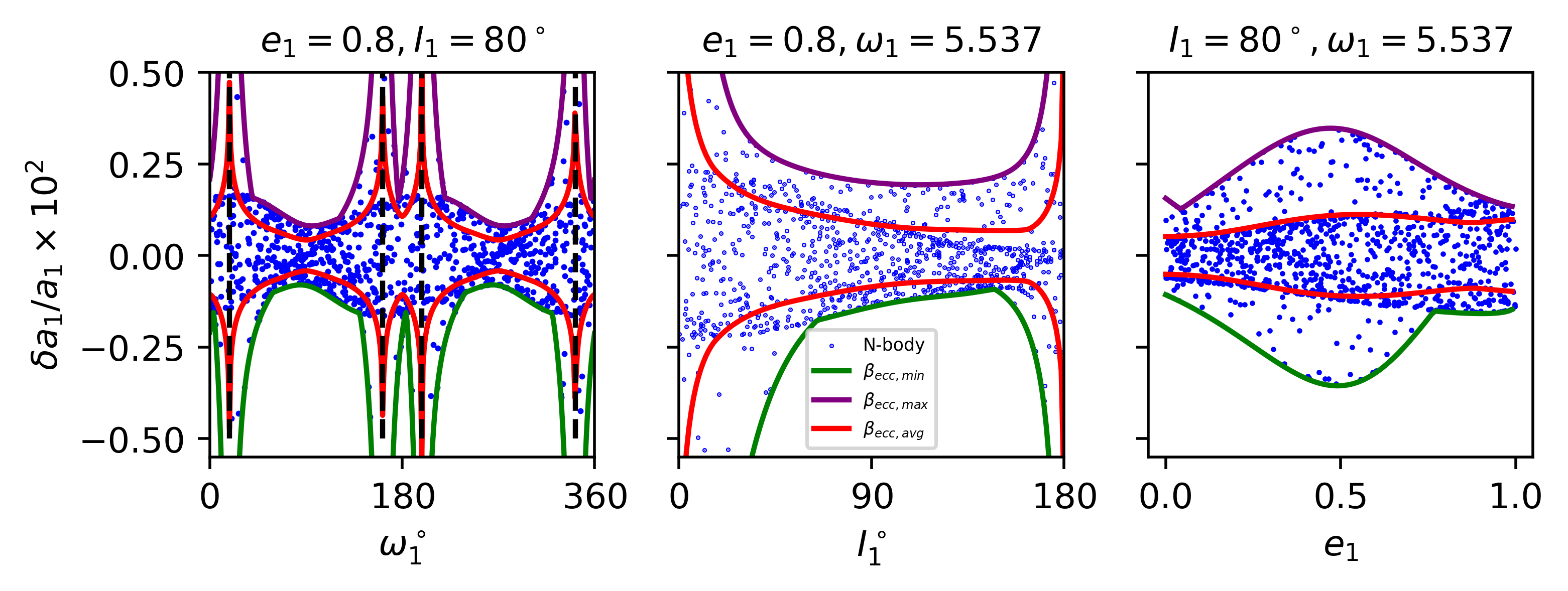}
	\caption{The left, middle, and right panels show the fractional change in the semi-major axis of the companion as a function of the argument of pericenter, inclination, and the eccentricity of the inner orbit. Results from N-body simulations are shown using the blue dots. The y-axis shows the fractional change in the semi-major axis of the inner orbit. The maximum ($\beta_{ecc,max}$), minimum ($\beta_{ecc,min}$) and average ($\beta_{ecc,avg}$) values of $\beta_{ecc}$, calculated using Eqn. \ref{eqn:dela1ecc}, are shown using the purple, green, and red lines respectively. We can see that on short timescales semi-analytical envelopes are in excellent agreement with N-body simulations. We use the following initial conditions for N-body simulations: $a_1 = 1 $ AU, $f_1=0$, $\Omega_1=0$, $a_2=1.5$ AU, $e_2=0$, $m_0 = 1 M_\odot, m_1=0$ and $m_2=10^{-3} M_\odot$. $f_2$ is chosen uniformly between 0 and $2\pi$. Other initial conditions are given at the top of the panel. N-body simulations are run for 1 orbital period i.e., $t_{max} =1$ year (blue dots) and 100 years (grey dots).}
	\label{fig:dela1e1}
\end{figure*}

We now compare the semi-analytical results from Eqn. \ref{eqn:dela1ecc} with N-body integration for a broad range of orbital elements. Figure \ref{fig:dela1e1} shows the fractional change in the semi-major axis of the inner orbit as a function of the argument of pericenter (left panel), inclination (middle panel), and the eccentricity (right panel) of the inner orbit. Results from N-body, which are run for 1 orbital period, are shown using the blue dots. The maximum and minimum values of $\beta_{ecc}$ (calculated semi-analytically using Eqn. \ref{eqn:dela1ecc}) are shown using the purple and the green lines respectively. We can see that the semi-analytical envelope agrees very well with N-body simulations. The semi-analytical average kick ($\pm \beta_{ecc,avg}$) is shown using the red lines. As expected, most of the N-body results lie within the bounds given by $\pm \beta_{ecc,avg}$. The results from N-body simulations run for 100 years are shown in grey. We can see that these results are not within the semi-analytical bounds. But we find that over longer timescales N-body results are within a factor of 2 of the semi-analytical bounds, as long as the orbital elements of the inner binary do not change significantly.

It should be noted that eccentric orbits in non-hierarchical systems can cross each other. The crossing condition is given by: $a_1(1-e_1^2)/(1\pm e_1\cos{\omega_1}) =a_2$. The left panel shows the crossing condition as the black dashed line. Similarly, in the middle panel, the inner orbit crosses the outer orbit when $I_1=0,180^\circ$. We can see that $\beta_{ecc}$ can become very large for crossing configurations. Meanwhile, none of the orbits shown in the right panel satisfy the crossing condition. As a result, $\beta_{ecc}$ is a relatively flat function of eccentricity. We find that crossing configurations are in general unstable over longer timescales \citep{zhangStabilityTimescaleNonhierarchical2023}.

\section{Secular stability limit}
It should be noted that the orbital elements of the inner orbit, and as a result $\beta_{ecc}$, can change significantly due to secular perturbations from the outer companion. From Figure \ref{fig:dela1e1}, we can see that $\beta_{ecc}$ has the strongest dependence on the inclination and the argument of the pericenter of the inner orbit. In general, for misaligned orbits, $\beta$ is maximized when the minimum distance between the inner and the outer orbits ($r_{min,orb} = \frac{a_1(1-e_1^2)}{(1 \pm e_1\cos{\omega_1}) )}-a_2$) is minimized. We denote the maximum value of $\beta_{ecc}$ over secular timescales as $\beta_{sec}$. To obtain a stability criterion that is valid for secular timescales we must use $\beta=\beta_{sec}$ in Equation \ref{eqn:stabcrit}.

To calculate $\beta_{sec}$, we must obtain the orbital elements of the inner orbit over secular timescales.  In general, secular equations of motion can be solved to obtain the time evolution of the orbital elements of the inner orbit. But, in the test particle approximation with a circular outer companion, the system has one degree of freedom, and a simpler method involving calculations of the contours of the secular Hamiltonian can be used.  Near the stability limit, the system is non-hierarchical, and an expansion of the Hamiltonian in the ratio of semi-major axes ($\alpha$) is inadequate. Hence, we obtain the secular Hamiltonian through direct numerical double averaging. 
\begin{equation} \label{eqn:hsecda}
	H_{sec,da} = \frac{1}{4\pi^2} \int_0^{2\pi} \int_0^{2\pi}  R  d\lambda_1 d\lambda_2 
\end{equation}
Multiple studies have used an approach for non-hierarchical systems \citep{gronchi_averaging_1998,beust_orbital_2016, saillenfest_non-resonant_2017,bhaskarMildlyHierarchicalTriple2021a}. The secular Hamiltonian along with the constant of motion $J_z = \sqrt{1-e_1^2}\cos{i_1}$ can then be used to obtain the trajectory of the inner binary.

\begin{figure*}
	\centering
	\includegraphics[scale=0.8]{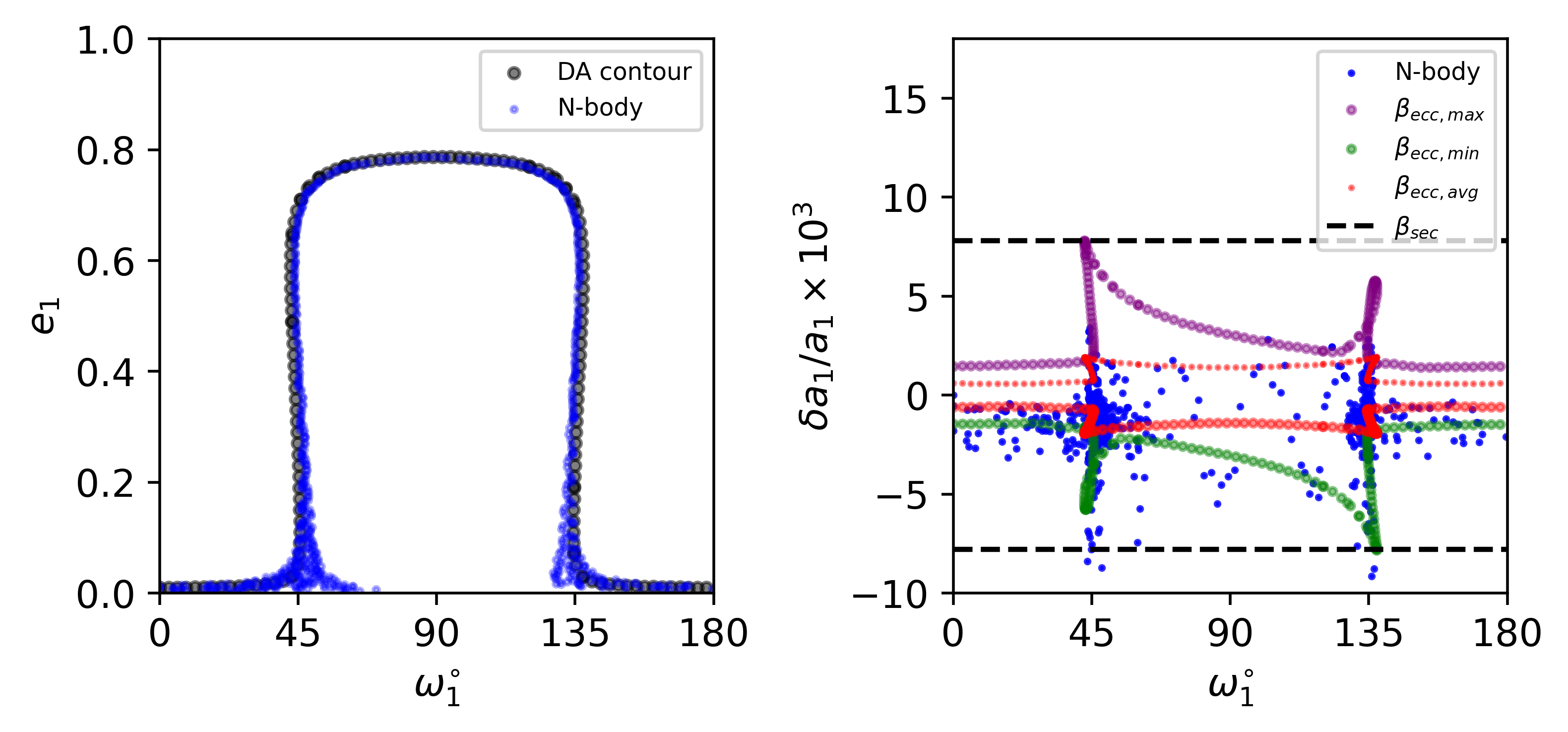}
	\caption{Calculation of $\beta_{sec}$, and comparison with N-body simulations. The x-axis shows the argument of the pericenter of the inner orbit. The y-axis shows the eccentricity in the left panel and the fractional change in the semi-major axis in the right panel. The contours of the numerically double-averaged secular Hamiltonian are shown in the left panel as black dots. N-body results are shown as blue dots. We can see that they agree with each other. The right panel shows $\beta_{ecc,min}, \beta_{ecc,max}$ and $\beta_{ecc,avg}$ for the points on the contour of the Hamiltonian. N-body results are again shown as blue dots. $\beta_{sec}$, which is defined as the maximum value of $\beta_{ecc,max}$, is shown using the horizontal black dashed lines. We can see that almost all of the blue dots lie inside  $\pm \beta_{sec}$. We use the following initial conditions to make this figure: $a_1 = 1 $ AU, $e_1=0.01$, $\omega_1=0$, $\Omega_1=0$, $a_2=1.4$ AU, $e_2=0.0$, $m_0 = 1 M_\odot, m_1=0$ and $m_2=10^{-3} M_\odot$.}
	\label{fig:secbetacalc}
\end{figure*}

Figure \ref{fig:secbetacalc} shows the calculation of $\beta_{sec}$ for an inner binary orbit initialized on a near-circular orbit. The left panel shows the contour of the numerically double-averaged Hamiltonian in black. The contour shows that the eccentricity of a near-circular orbit can be excited to 0.8 by secular perturbations. The orbital evolution as calculated by an N-body integration is shown in blue. We can see that the N-body results are in excellent agreement with the contours of the double-averaged Hamiltonian. $\beta_{ecc,max},  \beta_{ecc,min}$ and $\beta_{ecc,avg}$  are then calculated for the points on the contour of the Hamiltonian. These are shown in the right panel as purple, magenta, and green dots respectively. $\beta_{sec}$ is shown using the dashed black lines. The results from N-body simulations are shown in blue. We can see that most of the blue dots lie inside the bounds defined by the green and purple dots. There are a few dots outside these bounds. This is expected, as the semi-analytical bounds are exact only for one orbital period. Nevertheless, we can see that almost all the blue points lie inside $\pm \beta_{sec}$. Hence, we conclude that $\beta_{sec}$ serves as an excellent estimate of $\beta$ over secular timescales. 

\begin{figure}
	\centering
	\includegraphics[scale=0.5]{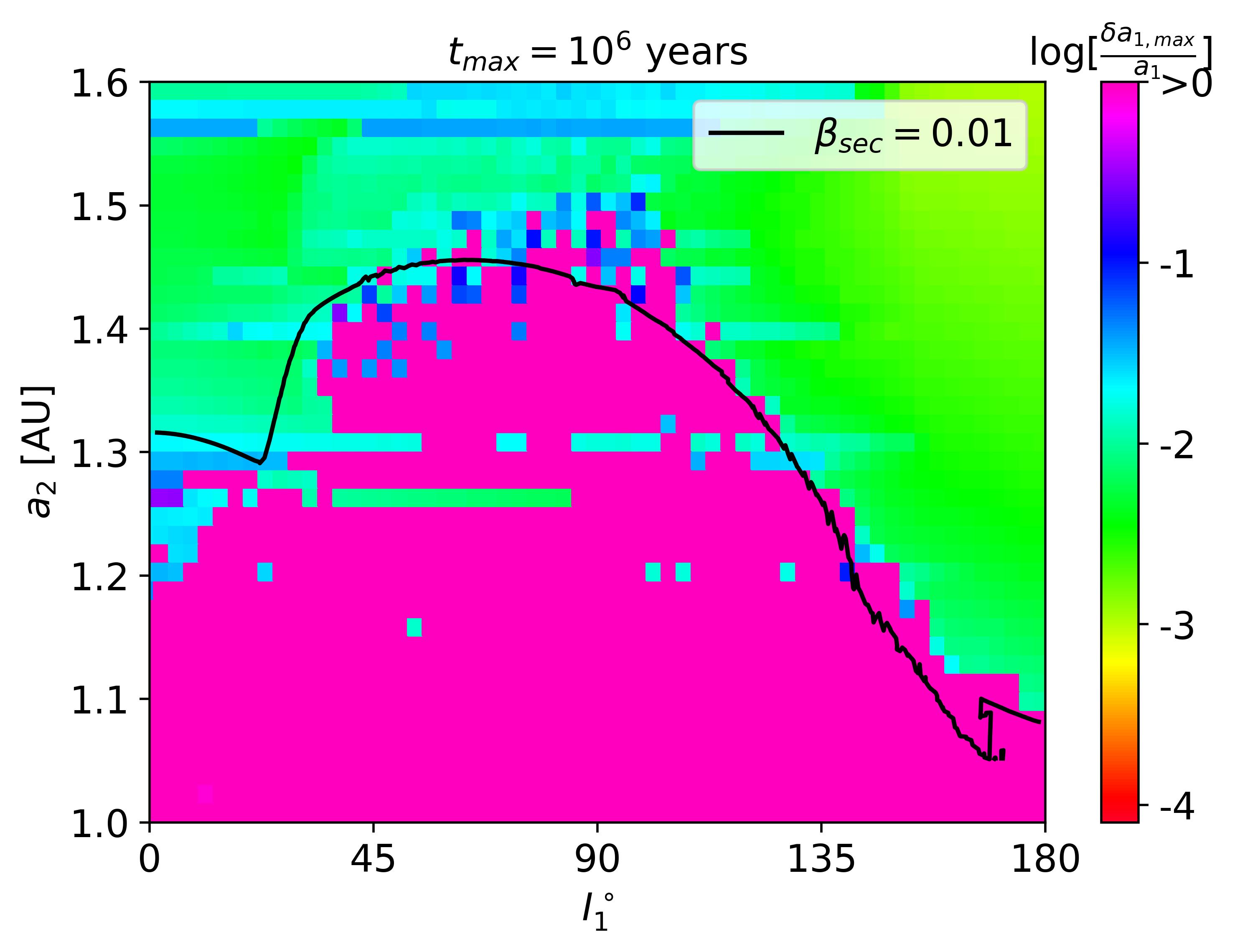}
	\caption{Comparison of semi-analytical stability limit $\beta_{sec}=0.01$ with N-body simulations. The color shows the maximum fractional change in the semi-major axis of the inner orbit as calculated by N-body simulations after $10^6$ (right panel) years of evolution. The black lines show the contour of $\beta_{sec}=0.01$. We can see that the semi-analytical criteria are in excellent agreement with N-body simulations. The more stable regions inside the instability, result from MMRs. We chose the following initial conditions for N-body simulations: $a_1 = 1$ AU, $e_1=0.01$, $e_2=0$, $m_0 = 1 M_\odot, m_1=10^{-6} M_\odot$ and $m_2=10^{-3} M_\odot$. Other angles including $\omega_1,\Omega_1, f_1$ and $f_2$ are chosen uniformly between 0 and $2\pi$.}
	\label{fig:nbodystabsecbeta}
\end{figure}

To find the secular stability criteria we calculate $\beta_{sec}$ for a wide range range of semi-major axis ratios and mutual inclinations (similar to the calculations shown in Fig. \ref{fig:a1inc1an}). In our calculations we focus on the contours of the secular Hamiltonian corresponding to initial conditions: $\omega_1=\pi/2$ and $e_1=0.01$. We numerically evaluate the expression $\beta_{sec}=\beta_{crit} =0.01$ to get the semi-analytical stability criteria. Figure \ref{fig:nbodystabsecbeta} compares the stability limit with N-body simulations. We can see that the stability limit is in excellent agreement with N-body simulations. The dependence of the stability limit on the initial mutual inclination is well reproduced by the contour $\beta_{sec}=0.01$.  A discontinuity can be seen in the contour at around $170^\circ$. This is because the inner binary transitions from librating to circulating orbits at this inclination. As a result, the system switches from a crossing configuration ($a_1(1+e_{max})>a_2$)  to a non-crossing configuration. 

It should be noted that the calculations shown in Figure \ref{fig:nbodystabsecbeta} only consider secular perturbations. Consequently, stable regions near 1.26 AU and 1.3 AU which correspond to the nominal locations of MMRs, are inconsistent with our stability criteria. In the following section, we discuss the importance of MMRs on the stability of two-planet systems.

\section{Stability of Planets in MMRs}
\label{sec:mmr}

\begin{figure}[h]
	\centering
	\includegraphics[width=1.0\linewidth,height=0.75\linewidth]{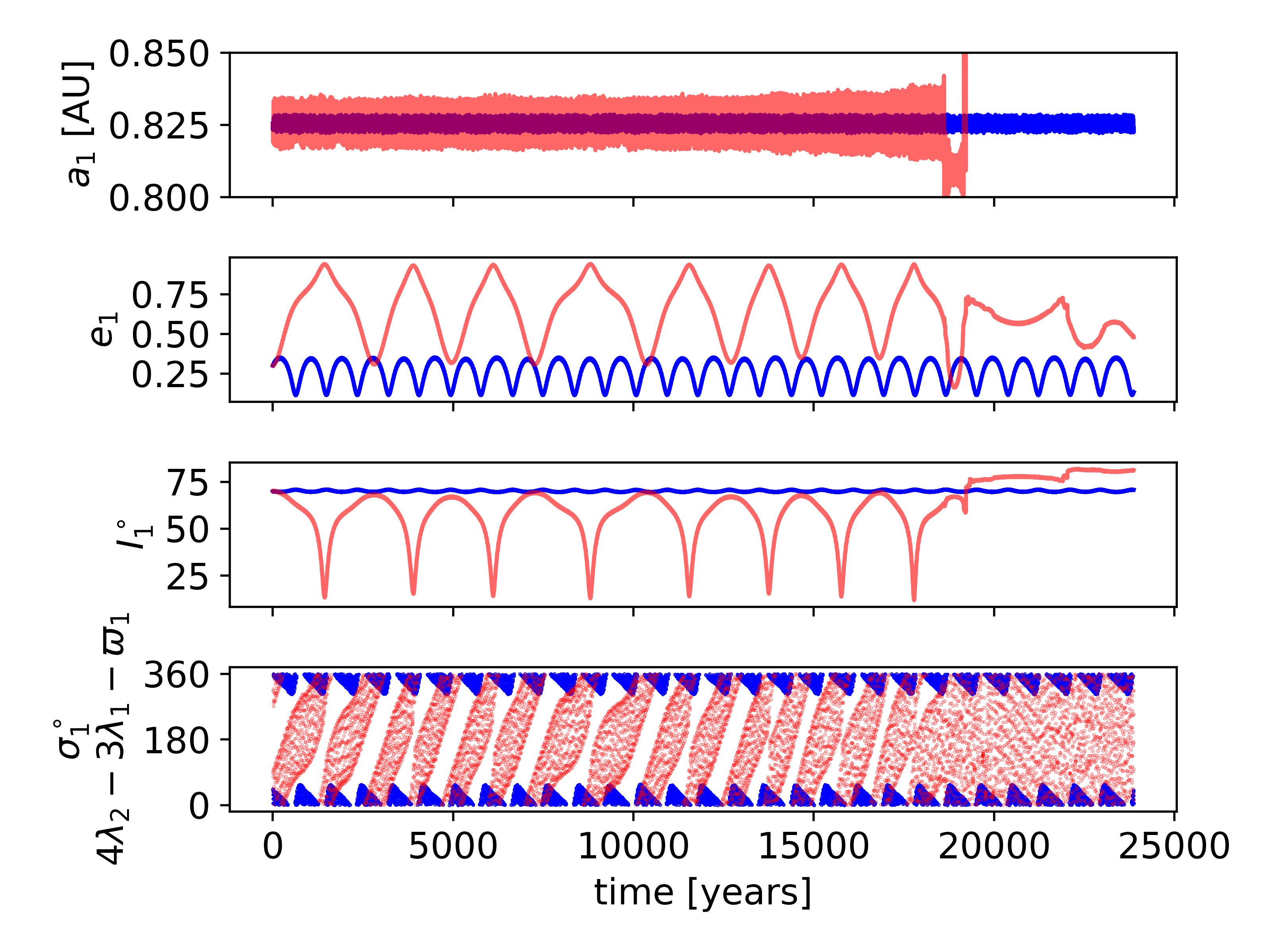}
	\caption{ Orbital evolution of misaligned inner orbit initialized at the nominal location of 4:3 mean motion resonance. The trajectories shown in red and blue differ only in the initial phase of the resonant angle $\sigma_1 = 4\lambda_2-3\lambda_1-\varpi_1$. We can see that while the blue trajectory librates around $\sigma_1 = 0$, the red trajectory is circulating.  In the red trajectory, we can see large variations in eccentricity and inclination similar to vZLK oscillations. The eccentricity excitation destabilized the inner orbit at around $1.8 \times 10^4$ years. These oscillations are suppressed in the blue trajectory, and the orbit remains stable. We use the following initial conditions to make this figure: $e_1=0.3$, $I_1=70^\circ$, $e_2=0$, $m_0 = 1 M_\odot, m_1=10^{-6} M_\odot$ and $m_2=10^{-3} M_\odot$.}
	\label{fig:nbdyMMRHE}
\end{figure}

%\begin{figure}
%	\centering
%	\includegraphics[width=1.0\linewidth,height=0.75\linewidth]{figures/nbodyRunMMRHE.png}
%	\caption{Orbital evolution of the inner planet misaligned with respect to the outer planet. The two planets are initialized at the nominal location of 2:1 mean motion resonance. The initial mutual inclination is 70 degrees. We can see large variations in  the eccentricity and inclination of the inner planet similar to Kozai-Lidov oscillations.}
%	\label{fig:nbdyMMRHE}
%\end{figure}

%\begin{figure}
%	\centering
%	\includegraphics[width=1.0\linewidth,height=0.75\linewidth]{figures/nbodyRunMMRLE.png}
%	\caption{Similar to Figure \ref{fig:nbdyMMRHE}, expect with a different initial value of the resonant angle. The secular evolution similar to Kozai-Lidov oscillations is suppressed in this run.}
%	\label{fig:nbdyMMRLE}
%\end{figure}
In our discussion so far we have focused on the importance of secular perturbations on the stability of two planet systems. Figure \ref{fig:stabNbody} shows that MMRs can also influence the stability. More specifically, we can see that some of the planets initialized at the nominal locations of 4:3, 7:5, and 3:2 MMRs are stable, even though planetary systems with $\alpha>0.66$ ($a_2<1.5$ AU in Figure \ref{fig:stabNbody}) are generally unstable. In this section, we will study in detail how MMRs affect the stability of a two-planet system.

It should be noted that not all trajectories initialized near the nominal locations of MMRs are stable (see Fig. \ref{fig:stabNbody}). The stability outcome depends on the phase of the resonance angle. This can be seen in Figure \ref{fig:nbdyMMRHE}, which shows the orbital evolution of two planets initialized at the nominal location of 4:3 MMR. The initial conditions for the planet shown in blue (red) are chosen such that the initial phase of the resonance angle $\sigma=4\lambda_2-3\lambda_1-\varpi_1$ is $0$ ($180^\circ$). Similar to vZLK oscillations, we can see that the eccentricity and the inclination of the planet shown in red go through large amplitude oscillations. The eccentricity is excited to values close to 1, which destabilizes the orbit at around $1.8 \times 10^4$ years (see top panel). Meanwhile, the eccentricity and the inclination oscillations are suppressed in the blue trajectory. The semi-major axis oscillations also have a smaller amplitude. As a result, the blue orbit remains stable for the entire duration of the simulation. The bottom panel shows that the blue (red) planet is in a librating (circulating) orbit. We can see that the eccentricity and the inclination oscillations of the inner orbit are suppressed when the resonance angle librates and the suppression of eccentricity excitation stabilizes the inner orbit.

% \begin{figure*}
% 	\centering
% 	\includegraphics[width=1.0\linewidth,height=0.35\linewidth]{stability_inc_dependence_tmaxall.png}
% 	\caption{Convergence of stability limit.}
% 	\label{fig:histmmrfo}
% \end{figure*}

% \begin{figure*}
% 	\centering
% 	\includegraphics[width=1.0\linewidth,height=0.35\linewidth]{stability_inc_dependence_mratall.png}
% 	\caption{Dependence on the mass of the inner planet.}
% 	\label{fig:histmmrfo}
% \end{figure*}

% \begin{figure}
% 	\centering
% 	\includegraphics[width=1.0\linewidth,height=0.8\linewidth]{stability_inc_dependence_tmax.png}
% 	\caption{Instability timescale.}
% 	\label{fig:histmmrfo}
% \end{figure}

\begin{figure*}
	\centering
	\includegraphics[width=0.8\linewidth,height=0.35\linewidth]{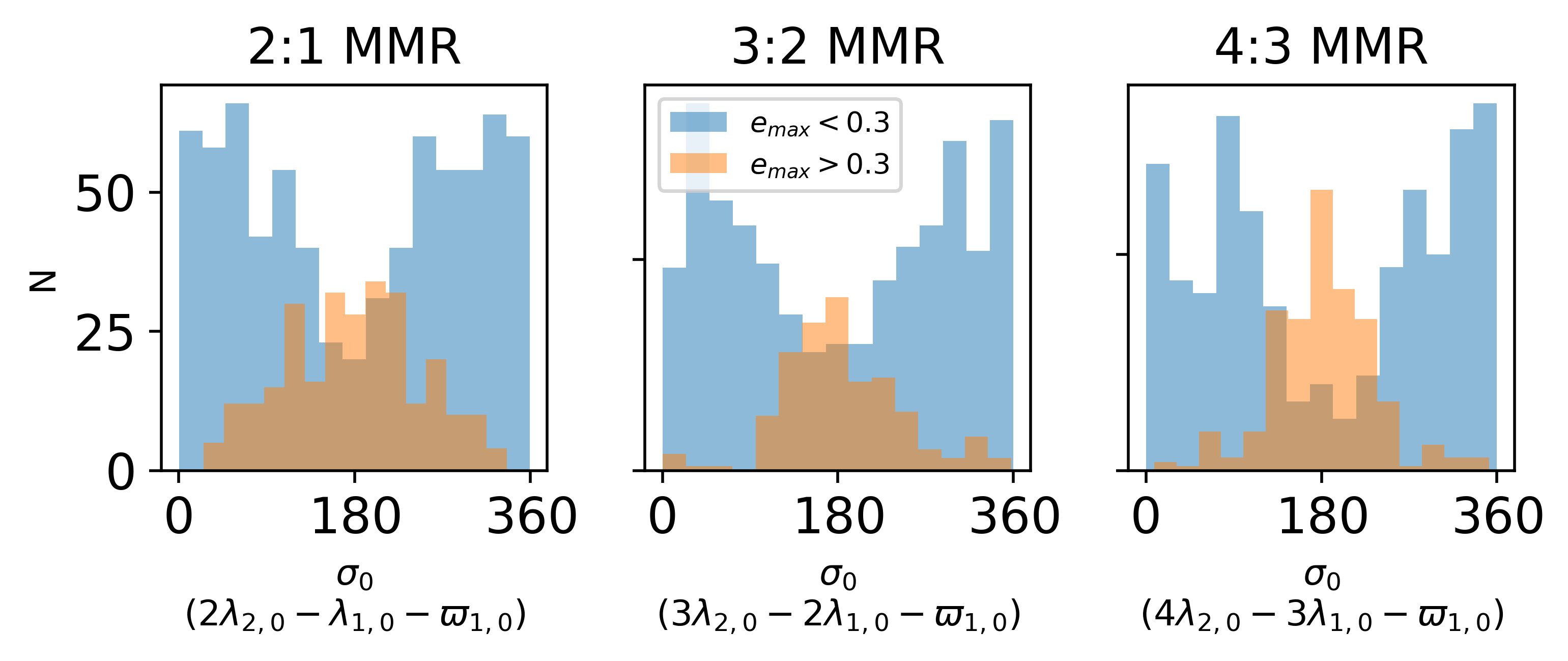}
	\caption{Dependence of the maximum eccentricity on the initial phase of resonance angle (from N-body simulations). The inner orbit is initialized at the nominal locations of first-order mean-motion resonances. The left, middle, and right panels show distributions for 2:1, 3:2, and 4:3 MMRs respectively. The distribution of the initial resonance angle for runs in which the maximum eccentricity is greater (lesser) than 0.3 is shown in orange (blue). We use the following initial conditions to make this figure: $a_1 = 1 $ AU, $e_1=0.1$, $I_1=70^\circ$, $e_2=0$, $m_0 = 1 M_\odot, m_1=10^{-6} M_\odot$ and $m_2=10^{-3} M_\odot$. The other planet is initialized at the nominal location of the MMR. Other orbital elements including $\omega_1,\Omega_1, f_1$ and $f_2$ are chosen uniformly between 0 and $2\pi$. }
	\label{fig:histmmrfo}
\end{figure*}

\begin{figure*}
	\centering
	\includegraphics[width=0.6\linewidth,height=0.5\linewidth]{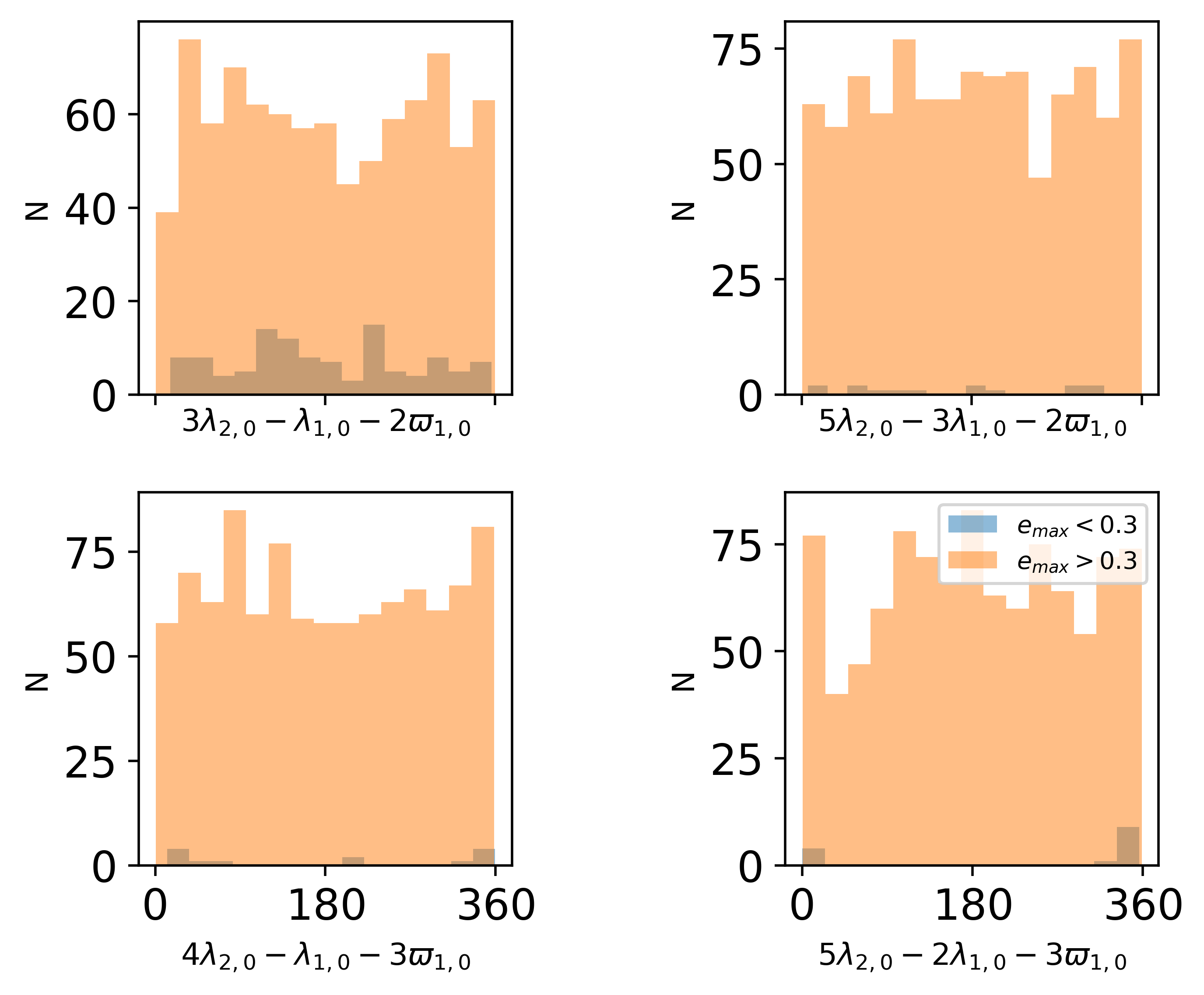}
	\caption{ Dependence of the maximum eccentricity on the initial phase of resonance angle (from N-body simulations). The runs are initialized at the nominal locations of second (top row) and third (bottom row) order mean-motion resonances. We show results for $3:1$ (top left), $5:3$ (top right), $4:1$ (bottom left) and $5:2$ (bottom right) MMRs. The distribution for runs in which maximum eccentricity is greater (lesser) than 0.3 is shown in orange (blue). We use the following initial conditions to make this figure: $a_1 = 1 $ AU, $e_1=0.1$, $I_1=70^\circ$, $e_2=0$, $m_0 = 1 M_\odot, m_1=10^{-6} M_\odot$ and $m_2=10^{-3} M_\odot$. The other planet is initialized at the nominal location of the MMR. Other orbital elements including $\omega_1,\Omega_1, f_1$ and $f_2$ are chosen uniformly between 0 and $2\pi$. }
	\label{fig:histmmrho}
\end{figure*}

We verified the results shown in Fig. \ref{fig:nbdyMMRHE} by running an ensemble of N-body simulations. In our simulations, we initialize the planets at the nominal location of $p:q$ MMRs. The initial resonance angle ($\sigma = p\lambda_2-q\lambda_1 -(p-q)\varpi_1$) is chosen uniformly between 0 and $2\pi$. The initial eccentricity of the inner planet is set to 0.01, and the initial mutual inclination between the planets is set to $60^\circ$. The simulations are run for a maximum of $t_{max}=10^5$ years. We record the maximum eccentricity ($e_{max}$) attained by the inner binary throughout the simulation.

Figure \ref{fig:histmmrfo} shows the distribution of the initial resonance angle for planets initialized at the nominal location of first-order mean motion resonances. The distribution for planets in which $e_{max} > 0.3$ ($e_{max}<0.3$) is shown in orange (blue). We can see that the distribution peaks at 0$^\circ$ and 360$^\circ$ ($\cos(\sigma) \approx 1$) for planets in which $e_{max}<0.3$. Meanwhile, the eccentricity is excited to large values for planets initialized near $\sigma \sim 180^\circ$. Comparing the different panels, we can see similar behavior for all the first-order MMRs. Figure \ref{fig:histmmrho} shows similar distributions for second and third-order MMRs. We can see that the secular eccentricity excitation is not suppressed for planets initialized at these resonances.  This is because these higher-order MMRs are weaker and have smaller resonance widths. %This is consistent with Fig. \ref{fig:stabNbody}, where stable inner planets are initialized at first order MMRs.

%More specifically, the coefficients of a specific resonant argument in the disturbing function is proportional to eccentricity raised to the order of the resonance. Meanwhile, the secular terms are proportional to $e^2$. Hence, coefficients of all resonant arguments other than the first order MMRs  are smaller than the secular terms.

 \begin{figure*}
 	\centering
 	\includegraphics[width=1.0\linewidth,height=0.26\linewidth]{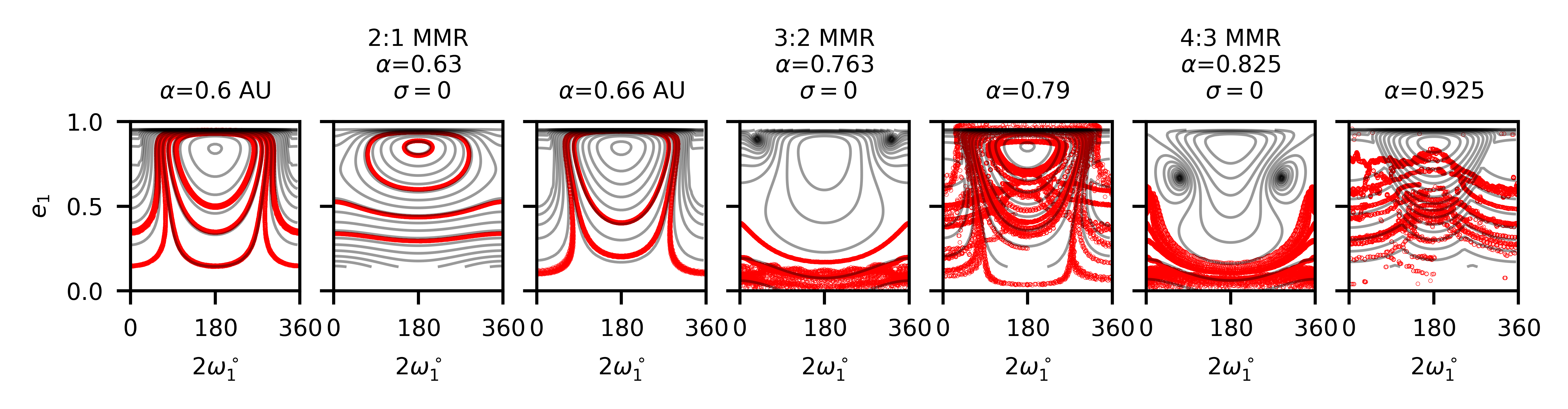}
 	\caption{Contours of the Hamiltonian for secular and resonant dynamics. The first, third, fifth, and last panels lie outside the resonance widths of MMRs. The contours of the Hamiltonian for these panels are calculated using numerical double averaging. The second, fourth, and sixth panels correspond to the locations of 2:1, 3:2, and 4:3 MMRs. A single averaged Hamiltonian is used to get the contours. In our calculations, we set the resonance angle to 0. The red dots show results from N-body simulations. We can see that the N-body integrations are consistent with contours, except when the $\alpha \geq 0.79$. By looking at the contours, it is clear that the eccentricity of near-circular orbits can be excited to large values when secular dynamics is important. MMRs suppress the eccentricity excitation, and as a result, stabilize the orbit.}
 	\label{fig:contmmr21}
\end{figure*}

We will now try to understand our N-body results using perturbation theory. Figures \ref{fig:contmmr21} show the contours of the Hamiltonian which dictates the long-term dynamics of the system (as solid black lines), along with results from direct N-body integrations (as the red dots). Outside the resonance widths of MMRs, secular dynamics are dominant. We use the secular Hamiltonian given by Eqn. \ref{eqn:hsecda} to follow the long-term evolution of the system. The contours of the secular Hamiltonian are shown in the first, third, fifth, and seventh panels of Figure \ref{fig:contmmr21} as black lines.

The dynamics of MMRs are usually studied by expanding the disturbing function in the eccentricity and the inclination of the binary. Laplace coefficients can be used in such expansions. Some studies avoid expansions in the inclination by utilizing two-dimensional Laplace coefficients. But here we need an approach that works for both arbitrary eccentricities and inclinations. Hence, we use a numerically single averaged disturbing function:\begin{equation}
	 H_{res,sa} = \frac{1}{2\pi}  \int_0^{2\pi} R (\lambda_1,\lambda_2,e_1,I_1,\omega_1,\Omega_1) \bigg \lvert_{\sigma}  d\lambda_1 
\end{equation} 
where the resonance angle ($\sigma$) is kept constant during the integration. This approach has also been used before in literature (e.g., \cite{gallardoSemianalyticalModelPlanetary2021}). The dynamics near MMRs are complicated by the additional degree of freedom associated with the resonant angle ($\sigma$) and the corresponding conjugate momentum. From Figures \ref{fig:nbdyMMRHE}, \ref{fig:histmmrfo} and \ref{fig:histmmrho} we can see that the eccentricity excitation is suppressed when the resonance angle librates around 0. Hence, to simplify the analysis, we set the resonance angle to 0 in our calculations of the contours of $H_{res,sa}$. This freezes the evolution of the degree of freedom associated with the resonance angle and helps us focus on the long timescale dynamics (see for example, \cite{kozaiSecularPerturbationsResonant1985} for a similar analysis). Similar to secular calculations, $J_z$ is kept constant to evaluate the contours of $H_{res,sa}$. The contours are shown in the second, fourth, and sixth panels of Figure \ref{fig:contmmr21} for 2:1, 3:2, and 4:3 MMRs respectively as the solid black lines.

We first look at the panels where secular perturbations are dominant (first, third, fifth, and seventh panels). Starting from near-circular orbits, we can see that the eccentricity of the inner orbit can be excited to large values ($\sim 1$). This is similar to vZLK oscillations. We can see that when the semi-major axis ratio is low ($\alpha < 0.76$), the N-body results are in excellent agreement with the contours of the Hamiltonian. At higher values of $\alpha$ (fifth and seventh panels), the N-body results do not follow one specific contour, and instead jump between different contours. This is due to the close encounters that are possible in this non-hierarchical system, causing substantial changes in the semi-major axis of the inner orbit.

Now looking at the panels where MMRs are important (second, fourth, and sixth panels), we can see that the N-body results are in excellent agreement with the contours of single averaged Hamiltonian. This is even though $\sigma$ is kept constant (=0). Comparing with the secular contours, we can see that the eccentricity of a near-circular orbit cannot be excited to 1 when MMRs are important. We hence conclude that the enhanced stability of the inner planet near MMRs (seen in Figure \ref{fig:stabNbody}) can be explained by the suppression of secular perturbations by MMRs.
% \begin{table*}
% 	\caption{List of systems in which hot Jupiters have an inner companion}
% 	\begin{tabular}{|c|c|c|c|c|c|c|c|c|c|}
% 		\hline	
% 		Citation & Name &$N_{comp}$ & $m_1$ $(M_\oplus)$ & $m_2$ $(M_J)$ & $a_1$ [AU]& $a_2$ [AU]& $m_1/m_2$ & 1/$\alpha$ & comments\\
% 		\hline
% 		\cite{lillo-boxTOI969LateKDwarf2023} &TOI-969&2& 8.89 & 11.8 & 0.02635 & 2.54 & 0.0024 &96.39 & $e_2=0.622$\\ 
% 		\hline
% 		\cite{canasKepler730HotJupiter2019} &Kepler-730&2&Earth mass&$<13$&0.03997 & 0.0694&$\sim 10^{-3}$&1.732  & masses unknown \\
% 		\hline
% 		\cite{shaTESSSpotsMinineptune2023} &TOI-2000&2&11.0&0.257&0.04271 & 0.0878&0.1346&2.06 & masses comparable \\
% 		\hline
% 		\cite{hordDiscoveryPlanetaryCompanion2022} &WASP-132&2&37.35&0.41&0.0182&0.067&0.286&3.68132& masses comparable \\
% 		\hline
% 		\cite{kaneDarkPlanetsWASP472020} &WASP-47&4&9.0&1.14&0.017&0.052&0.024&3.05882& 4-planet system \\
% 		\hline
% 		\cite{korthTOI1130PhotodynamicalAnalysis2023} &TOI-1130&2&12.9&0.974&0.04394&0.07098&0.0416&1.615& 2:1 MMR \\
% 		\hline
% 	\end{tabular}
% \end{table*}
\section{Discussion}
\label{sec:discuss}

\subsection{Validity of semi-analytical results over longer timescales}
\label{subsec:ltsc}
We saw in section \ref{sec:eccstab} that our N-body simulations run for $10^6$ years are consistent with the semi-analytical stability criteria (see Fig. \ref{fig:nbodystabsecbeta}). But, it should be noted that the stability limit can vary significantly over time (see Fig. \ref{fig:compstabnbd1e2}). Also, since most of the observed planetary systems are billions of years old, it is important to establish the validity of our stability limit over longer time periods. 

\begin{figure*}
	\centering
	\includegraphics[scale=1.0]{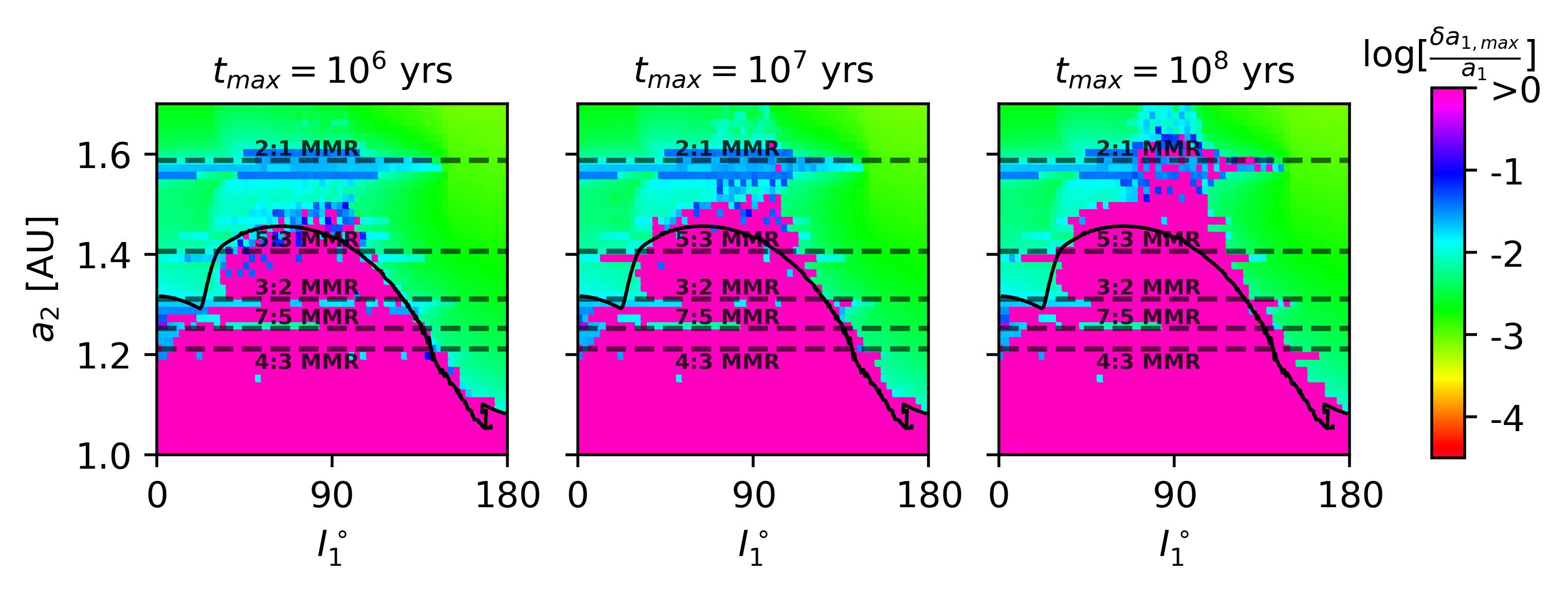}
	\caption{
 Comparison of the semi-analytical stability criteria ($\beta_{sec}=0.01$) with N-body simulations. The color shows the maximum fractional change in the semi-major axis of the inner orbit as calculated by N-body simulations after $10^6$ (left panel), $10^7$ (middle panel), and $10^8$ (right panel) years of evolution. The solid black lines show the contour of $\beta_{sec}=0.01$ from Figure \ref{fig:nbodystabsecbeta}. We can see that the contour is consistent with N-body simulations. We choose the following initial conditions for N-body simulations: $a_1 = 1$ AU, $e_1=0.01$, $e_2=0$, $m_0 = 1 M_\odot, m_1=10^{-6} M_\odot$ and $m_2=10^{-3} M_\odot$. Other angles including $\omega_1,\Omega_1, f_1$ and $f_2$ are chosen uniformly between 0 and $2\pi$.
 }
	\label{fig:stabcomptmax}
\end{figure*}

Figure \ref{fig:stabcomptmax} compares our semi-analytical stability limit with N-body simulations after $10^6$, $10^7$, and $10^8$ years of evolution. The dependence of the stability limit on the initial mutual inclination is still well modeled by the semi-analytical stability criteria. Comparing the different panels we can see that planets initialized near the nominal location of MMRs become unstable over longer timescales. For instance, the planets near 3:2 MMR which were stable at $10^6$ years, become unstable by $10^8$ years. Also, planets near 2:1 MMR become more unstable after $10^8$ years. Over longer timescales, planets even outside MMR become unstable. The instability is driven by the chaotic evolution of non-hierarchical systems. One of the possible ways to address this issue is to decrease the value of $\beta_{crit}$ with time. For the timescales, we are interested in ($\lesssim 10^9$ years), $\beta_{crit}$ would vary by less than an order of magnitude. Hence, for simplicity, we choose $\beta_{crit}=0.01$ for all timescales.

\subsection{Comparison with observations}
\label{subsec:compobs}
\begin{table*}
	\caption{List of systems in which hot Jupiters/Saturns have a low-mass inner companion.}
    \label{tab:loobs}
	\begin{tabular}{|c|c|c|c|c|c|c|c|c|c|c|}
		\hline	
		Citation & Name &$N_{planets}$ & $m_1$ $(M_\oplus)$ & $m_2$ $(M_J)$ & $m_0 (M_\odot)$& $a_1$ [AU]& $a_2$ [AU]& $m_1/m_2$ & $a_2/a_1$ & comments\\
		\hline
		\cite{lillo-boxTOI969LateKDwarf2023} &TOI-969&2& 9.1 & 11.3 & 0.735 &0.02636 & 2.52 & 0.0025 &95.6 & $e_2=0.622$\\ 
		\hline
		\cite{canasKepler730HotJupiter2019} &Kepler-730&2& 3-4 &$<13$& 1.047 &0.03997 & 0.0694&$\sim 10^{-3}$&1.732  & masses unknown \\
		\hline
		\cite{shaTESSSpotsMinineptune2023} &TOI-2000&2&11.0&0.257 & 1.083 &0.04271 &0.0878&0.1346&2.06 & masses comparable \\
		\hline
		\cite{hordDiscoveryPlanetaryCompanion2022} &WASP-132&2&$<37.35$&0.41 & 0.781 &0.0182&0.067&0.286&3.68132& masses comparable \\
		\hline
		\cite{kaneDarkPlanetsWASP472020} &WASP-47&4&9.0&1.14& 1.11 &0.017&0.052&0.024&3.05882& 4-planet system \\
		\hline
		\cite{korthTOI1130PhotodynamicalAnalysis2023} &TOI-1130&2&12.9 &0.974&0.712&0.04394&0.07098&0.0416&1.615& 2:1 MMR \\
		\hline
	\end{tabular}
\end{table*}

Close to 5600 exoplanets have been discovered in the past couple of decades. Many exoplanets have planetary companions. To compare with our analytical and semi-analytical results we focus on two planet systems (excep WASP-47 which contains 4 planets). More than 300 two-planet systems have been observed so far. In most of these systems, the ratio of the mass of the inner planet and the outer planet is close to 1. Consequently, our current results cannot be directly applied to most of these systems. 

More recently, multiple low-mass planets have been detected inside the orbits of hot Jupiters/Saturns. These planets are discovered mainly through transit surveys. We focus on these planets because they are more likely to be near the stability limit. Table \ref{tab:loobs} shows the list of these planets. We note that the inclination, mass, and age-dependent stability criteria can sometimes be used to constrain the orbital and physical properties of the planetary systems. We discuss some of these aspects below but postpone a more detailed discussion to a future study which would also include stability criteria and results for systems with comparable planetary masses, not explored here, and for which a far larger observational dataset is available.   

{\bf TOI-969} contains a mini-Neptune and an eccentric cold Jupiter. The system is hierarchical, with a semi-major axis ratio of around 100. The inclination of the Jupiter is not known. It is apparent that TOI-969 is quite different from the setup we use in this study, and hence a direct comparison with our work is not appropriate. Nevertheless, we can see that the hierarchical configuration of the system would help its long-term stability. 

The {\bf Kepler-730} planetary system was discovered using the transit method. It contains an earth-mass planet and a hot Jupiter in a relatively compact near-co-planar configuration. The masses of the planets are unknown. Assuming the upper bound on the mass of the companion  $m_2=13 M_{jup}$, we find $\beta_{circ,cp} =0.033$ (using Eqn. \ref{eqn:dela1aneqn}). We can see that such a high-mass companion would destabilize the inner planet.  From our stability criteria, we find that the inner planet is stable only when $m_2<4 M_{jup}$.

{\bf TOI-2000} contains a mini-Neptune and a hot Saturn. Both of these planets are on circular orbits and in a near co-planar configuration. The masses of the planets in this system are comparable ($m_2/m_1 \sim 0.135$).  While our analytical results cannot be directly applied to this system, we find that the system has a low value of $\beta_{circ,cp} (=2.8 \times 10^{-3}$).  

The transit method was also used to detect a super-Earth and a hot Jupiter in {\bf WASP-132}. The mass of the inner planet is not well constrained. The lower limit on the mass ratios is high ($m_2/m_1 > 0.2$). From Eqn. \ref{eqn:dela1aneqn} we get $\beta_{circ,cp} =4.65 \times 10^{-5}$. Hence, the long-term stability of the system is not ruled out by our stability criteria.  

{\bf WASP-47} contains 4 planets, with the inner three planets on near co-planar circular orbits.  The innermost planet is a super-Earth, and its nearest companion is a hot Jupiter. In this system, interactions between the inner planet and Jupiter’s outer companion (a super-Earth at $0.087$ AU) could be important for the long-term stability of the system. Interactions with a second outer companion are not included in this study. Nevertheless, we find $\beta_{circ,cp} = 2.5 \times 10^{-4}$, which does not rule out the long-term stability of the system. 

{\bf TOI-1130} contains a hot Jupiter and a super-Earth in a near co-planar configuration. The planets are near 2:1 MMR. We can see from Fig. \ref{fig:stabcomptmax} that a co-planar system near 2:1 MMR can remain stable for hundreds of secular timescales. This is consistent with the 3-5 billion year age of the system.

So far no misaligned planetary systems near the stability limit have been detected. A more rigorous verification of our stability criteria will hence require more observations.

\subsection{Dependence on the mass of the inner planet}
\label{subsec:mdep}
In our analytical and semi-analytical calculations, we have focused on systems in which the inner planet is much less massive than the outer planet. In practice, we found that our results are consistent with N-body simulations as long as the mass ratio of the inner and the outer planets ($m_1/m_2$) is less than $10^{-3}$. At higher mass ratios, the orbit of the outer planet is not constant, and the long-term evolution of the planets is coupled with each other. In general, planetary systems are more unstable at higher mass ratios. N-body simulations show that planetary systems with mass ratios of 0.1 and 1 become unstable in $10^6$ years when $\alpha=a_1/a_2$ is less than $2/5$ and 1/3 respectively. Similar to the test particle case, retrograde configurations are more stable than prograde configurations. Misaligned planetary systems are in general more unstable. Also, planets are more easily destabilized when initialized near the nominal locations of MMRs. More detailed stability analysis of planetary systems with comparable masses is left to a future study.

\section{Conclusions}
\label{sec:conc}
In this paper, we study the stability of mutually inclined two-planet systems. More specifically, we focus on planetary systems containing a Jupiter mass planet and a much less massive inner planet. Using an ensemble of N-body simulations, we deduce the parameter space where the orbit of the inner planet is stable. Our goal is to determine the critical mutual separation beyond which the inner orbit is stable (stability limit). In particular, we investigate the dependence of the stability limit on the mutual inclination between the two planets (see Fig. \ref{fig:stabNbody}). We find that retrograde orbits are more stable than prograde orbits. At large mutual inclinations ($40^\circ<I_1<140^\circ$), the inner orbit is destabilized by the eccentricity excitation induced by secular perturbations from the outer companion. Consequently, the stability limit occurs at larger separations for $40^\circ<I_1<150^\circ$. In addition to secular interactions, MMRs can also affect the stability of two body systems. Under certain conditions, MMRs can stabilize the orbits of the inner planet, by quenching the secular evolution effects. In contrast, stability criteria previously derived in literature, not accounting for secular effects and/or MMRs cannot fully account for the dependence of the stability limit on the mutual inclination between the planets (see Fig. \ref{fig:compstablit}). 

We use perturbation theory to derive a semi-analytical stability criteria which is applicable for arbitrary inclinations. To derive the stability criteria, we calculate the characteristic fractional change in the semi-major axis of the inner planet caused by perturbations from the outer companion ($\beta =\delta a_1/a_1$). We initially focus on circular inner orbits (Section \ref{sec:circstab}). In our derivation, we first ignore secular changes to the orbital elements of the inner orbit. An analytical expression for $\beta_{circ} (= \beta \vert_{e_1=0})$ is derived for $I_1=0$ and $I_1=180^\circ$ (Eqn. \ref{eqn:dela1aneqn}). For arbitrary inclinations, we evaluate a semi-analytic expression. We show that our analytical and semi-analytical results agree with direct N-body integrations as long as the secular changes in the orbital elements of the inner orbit are negligible (see Figures \ref{fig:compproan}, \ref{fig:comparbincan}). In our setup, this condition is satisfied for timescales less than 100 yrs. 

We then derive stability criteria by setting a threshold on $\beta$ i.e., the inner orbit is assumed to be stable if $\beta< \beta_{crit}$.  We show that the circular stability criteria ($\beta_{circ}< \beta_{crit}$) agree with direct N-body integrations on short timescales ($<$100 years). On secular timescales ($>10^4$ yrs), the eccentricity of misaligned inner orbits is excited to large values ($>0.3$), and the circular stability criteria are inadequate. Meanwhile, for near co-planar orbits, secular effects don't significantly change the eccentricity, and the criteria is valid for longer timescales (Fig. \ref{fig:compstabnbd1e2}). 

We generalize our calculation of $\beta$ to eccentric orbits ($\beta_{ecc}$) in section \ref{sec:eccstab}. We derive a semi-analytical expression for the change in the semi-major axis of the inner orbit over one orbital period (Eqn. \ref{eqn:dela1ecc}). We show that the semi-analytical results agree with N-body simulations for a wide range of orbital elements  (Figures \ref{fig:a1f1traj} and \ref{fig:dela1e1}). Over multiple orbital periods, semi-analytical results are still within a factor of 2 of N-body results. 

Over secular timescales, the orbital elements of the inner planet can change significantly. As a result, the $\beta_{ecc}$ also changes over secular timescales ($\beta_{ecc} \rightarrow \beta_{ecc}(t)$). To deduce the secular stability limit, we calculate the maximum value of $\beta_{ecc} (t)$  over secular timescales. Using contours of the double-averaged Hamiltonian, we calculate $\beta_{sec} = max(\beta_{ecc})$ for a wide range of semi-major axis ratios and mutual inclinations. We show that the semi-analytical expression $\beta_{sec}<0.01$ serves as an excellent stability criterion which is valid for a wide range of mutual inclinations.

The stability limit derived in section \ref{sec:eccstab} only takes secular perturbations into account. Figure \ref{fig:stabNbody} shows that MMRs can also affect the stability of two planet systems. Hence, we study the long-term dynamics induced by MMRs in section \ref{sec:mmr}. We find that the eccentricity excitation caused by secular perturbations is suppressed by first-order MMRs (see Figures \ref{fig:histmmrfo}, \ref{fig:histmmrho}). By drawing contours of the Hamiltonian, we show that the eccentricity of near-circular orbits initialized at the nominal location of first-order MMRs orbits can remain low over long timescales (see Fig. \ref{fig:contmmr21}). The suppression of the secular eccentricity excitation stabilizes the inner orbit (see Fig. \ref{fig:nbdyMMRHE}). 

Finally, we discuss the applicability of our results to systems with massive inner planets, consistency of our analysis with N-body simulations over different timescales, and comparison of our results with observed exosystems in section \ref{sec:discuss}.

\begin{acknowledgments}
The authors thank Billy Quarles, Gongjie Li, Douglas Lin and Cristobal Petrovich for useful discussions. HBP acknowledges support from the Minerva Center for Life under Extreme Planetary Conditions.
\end{acknowledgments}

\bibliography{ref}{}
\bibliographystyle{aasjournal}
\end{document}